\documentclass[aps,twocolumn,showpacs,floatfix,amssymb]{revtex4}
\usepackage{graphicx}
\begin{document}

\title{Dynamic  spin susceptibility in the $t$-$J$ model}
\author{ A.A. Vladimirov$^{a}$, D. Ihle$^{b}$, and N. M. Plakida$^{a,c}$ }
\affiliation{ $^a$Joint Institute for Nuclear Research, 141980 Dubna, Russia\\
$^{b}$ Institut f\"{u}r Theoretische Physik, Universit\"{a}t
Leipzig,
 D-04109, Leipzig, Germany \\
 $^{c}$Max-Planck-Institut f\"{u}r Physik komplexer Systeme,  D-01187,   Dresden,
Germany}

\date{\today}

\begin{abstract}
A  relaxation-function theory for the dynamic  spin
susceptibility in the $t$--$J$ model is presented. By a
sum-rule-conserving generalized mean-field approximation (GMFA),
the two-spin correlation functions of arbitrary range, the
staggered magnetization, the uniform static susceptibility, and
the antiferromagnetic correlation length are calculated in a wide
region of  hole doping and temperaturs. A good agreement with
available exact diagonalization (ED) data is found. The
correlation length is in reasonable agreement with
neutron-scattering experiments on
La$_{2-\delta}$Sr$_\delta$CuO$_4$. Going beyond the GMFA, the
self-energy is calculated in the mode-coupling approximation. The
spin dynamics at arbitrary frequencies and wave vectors is
studied for various  temperatures and  hole doping. At low doping
a spin-wave-type behavior is found  as in the Heisenberg model,
while at higher doping a strong damping caused by hole hopping
occurs, and a relaxation-type spin dynamics is observed in
agreement with the ED results. The local  spin susceptibility  and
its $\omega/T$ scaling behavior are calculated in a reasonable
agreement with experimental and ED data.
\end{abstract}


\maketitle

\section{Introduction}

It is generally believed that charge-carrier interaction with
spin fluctuations in the cuprate high-temperature superconductors
is the origin of their anomalous normal state properties and may
be responsible for the superconducting transition (see, e.g.,
Ref.~\cite{Chubukov02}).  Inelastic neutron scattering
experiments have revealed quite a complicated behavior of the
spin-fluctuation spectra in cuprates.~\cite{Kastner98,Bourges98}
Therefore, studies of spin fluctuations in these materials are
essential in elucidating the nature of high-temperature
superconductivity. Two limiting cases can be well described. In
the undoped insulating phase the quasi-two-dimensional Heisenberg
model for localized spins gives a reasonable description of the
spin-fluctuation spectra (see, e.g., Ref.~\cite{Manousakis91}),
while in the overdoped region the random phase approximation
(RPA) for weakly correlated itinerant electrons can be applied
(see, e.g., Ref.~\cite{Manske04}).
\par
However, the region  of light and optimal doping (the so-called
``pseudogap phase''), where localized spins on copper sites
strongly interact with correlated  charge carriers is much more
difficult to study. This region should be treated within  a model
of strongly correlated electrons like the Hubbard model
~\cite{Hubbard65} or the $t$–-$J$ model.\cite{Anderson87} The
charge-carrier motion in the $t$–-$J$ model is described by the
Hubbard projection operators, whose commutation relations are
more complicated than those of Fermi or Bose operators. Various
approaches have been used to study the spin dynamics in the
$t$–-$J$ model (for a review see, e.g., Refs.~\cite{Izyumov97}
and ~\cite{Plakida06}). In particular, in the slave boson or
fermion methods,  a local constraint prohibiting a double
occupancy of any quantum state is difficult to treat rigorously.
An application of a  special diagram technique for Hubbard
operators to the $t$--$J$ model results in a complicated
analytical expression for the dynamic spin susceptibility
(DSS).~\cite{Izyumov90} Studies of finite clusters by numerical
methods were important in elucidating static and dynamic spin
interactions, though they have limited energy and momentum
resolutions (see, e.g., Refs.~\cite{Dagotto94,Jaklic00,Eder95}).
\par
To overcome this complexity, we apply the projection Mori-type
technique elaborated for the two-time thermodynamic Green's
function (GF).~\cite{Zubarev60,Plakida73,Tserkovnikov81} In this
method an exact representation for the self-energy (or
polarization operator) can be derived which, when evaluated in
the mode-coupling approximation, yields  physically reasonable
results even for strongly interacting systems. As our calculations
have shown, the decoupling of the  correlation function of
currents, i.e., $\,( d S^+_{\bf q}/ dt )\, $, in
Ref.~\cite{Jackeli98} is insufficient for obtaining reasonable
results. Therefore, in the present paper,  in studying the DSS
$\chi({\bf q}, \omega)= - \langle\langle S^+_{\bf q}|S^-_{-\bf
q}\rangle\rangle_{\omega}$ (written in terms of the
GF~\cite{Zubarev60}) the mode-coupling approximation in the
paramagnetic phase is  applied to the correlation functions of
the forces, i.e., $\,(d^2 S^+_{\bf q}/ dt^2)\,
$.~\cite{Tserkovnikov81}  A similar approach based on the Mori
projection technique for the single-particle electron GF and spin
GF has been used in
Refs.~\cite{Sherman03,Sherman04,Sega03,Prelovsek04}. The magnetic
resonance mode observed in the superconducting state was studied
within the memory-function approach in
Refs.~\cite{Sherman03a,Sega03,Sega06,Prelovsek06}.
\par
In this paper  we use  the spin-rotation-invariant
relaxation-function theory for the DSS in the $t$--$J$ model
derived by us in Ref.~\cite{Vladimirov05} to calculate the static
properties in the generalized mean-field approximation (GMFA)
similarly to Ref.~\cite{Winterfeldt98} and the dynamic
spin-fluctuation spectra using the mode-coupling approximation
for the force-force correlation functions. Thereby, we capture
both the local and itinerant character of charge carriers in a
consistent way. In calculating the static properties, in
particular the static susceptibility and spin-excitation
spectrum, we pay particular attention to a proper description of
antiferromagnetic (AF) short-range order (SRO) and its
implications on the spin dynamics. For the undoped case described
by the Heisenberg model our results are similar to those in
Refs.~\cite{Winterfeldt97} and ~\cite{Winterfeldt99}. For a
finite doping our theory yields a reasonable agreement with
available exact diagonalization (ED) data and neutron scattering
experiments.
\par
The paper is organized as follows. In the next section the
$t$--$J$ model is formulated in terms of the Hubbard operators,
and basic formulas for the static spin susceptibility and the
self-energy within the relaxation-function
theory~\cite{Vladimirov05} are presented. Numerical results for
the static properties and spin-fluctuation spectra are given in
Sec.~III, where their temperature and doping dependences are
analyzed. The conclusion is given in Sec.~IV. Details of the
calculations are discussed in the Appendix.

\section{Relaxation-function theory}

\subsection{Basic formulas}

We start with the $t$--$J$ model on the square lattice,
\begin{eqnarray}
H &=& - \sum_{i \neq j,\sigma}t_{ij}X_{i}^{\sigma
0}X_{j}^{0\sigma}
 - \mu \sum_{i \sigma} X_{i}^{\sigma \sigma}
\nonumber \\
 &  + &\frac{1}{4} \sum_{i \neq j,\sigma} J_{ij}
\left(X_i^{\sigma\bar{\sigma}}X_j^{\bar{\sigma}\sigma}  -
   X_i^{\sigma\sigma}X_j^{\bar{\sigma}\bar{\sigma}}\right),
\label{b1}
\end{eqnarray}
which is written in terms of the Hubbard operators (HOs)
\cite{Hubbard65} $\,X_{i}^{\alpha\beta}=|i,\alpha\rangle\langle
i,\beta| \,$ for three possible states at a lattice site $i$: for
an empty site $|i,\alpha\rangle=|i,0\rangle$ and for a singly
occupied site $|i,\alpha\rangle=|i,\sigma\rangle$ with spin
$\sigma(1/2)$ ($\sigma = \pm  , \; \bar{\sigma} = - \sigma$). The
HOs obey the multiplication rule $\, X_{i}^{\alpha\beta}
X_{i}^{\beta \gamma} = X_{i}^{\alpha\gamma} \,$ and  the
completeness relation $ X_{i}^{00} + \sum_{\sigma}
X_{i}^{\sigma\sigma} = 1$, which preserves rigorously the
constraint of no double-occupancy of any lattice site. The spin
and number operators of the model  are given by $ S_{i}^{\sigma}=
X_{i}^{\sigma\bar{\sigma}}, \quad
S_{i}^{z}=(1/2)\sum_{\sigma}\sigma X_{i}^{\sigma\sigma}$, and
$n_{i}=\sum_{\sigma}X_{i}^{\sigma\sigma} $. The chemical
potential $\mu$ is  determined from the equation for the average
electron density   $ n = \sum_{ \sigma} \langle X_{i}^{\sigma
\sigma} \rangle = 1- \delta  $ where $\delta = \langle X_{i}^{00}
\rangle$ is the hole concentration.
\par
In Ref.~\cite{Vladimirov05} we have derived the general
expression for the DSS $\chi({\bf q}, \omega)= - \langle\langle
S^+_{\bf q}|S^-_{-\bf q}\rangle\rangle_{\omega}$,
\begin{equation}
\chi({\bf q}, \omega)= \chi_{\bf q}\, \frac{\omega_{\bf q}^2} {
\omega_{\bf q}^2 +\omega \, \Sigma({\bf q},\omega) - \omega^2} \,
.
 \label{b2}
\end{equation}
The static spin susceptibility $\chi_{\bf q}\,$ is related to a
generalized mean-field spin-excitation spectrum $\omega_{\bf q}$
by the equation
\begin{equation}
\chi_{\bf q} = ({S}_{\bf q}^{+},S_{-{\bf q}}^{-}) =  m({\bf q})
/\omega_{\bf q}^2
 \label{b3}
\end{equation}
with $m({\bf q})=\langle [i\dot{S}^{+}_{\bf q}, S_{-\bf
q}^{-}]\rangle$.  Here, the Kubo-Mori scalar product is defined
as (see, e.g.,~Ref.~\cite{Tserkovnikov81})
\begin{equation}
(A(t), B) =\int_{0}^{\beta}d\lambda \langle A(t-i\lambda) B
\rangle,
  \quad \beta = 1/k_{\rm B} T  .
 \label{b4a}
\end{equation}
 The self-energy is given by~\cite{Vladimirov05}
\begin{equation}
\Sigma({\bf q},\omega)=\frac{1}{m({\bf q})}(( - \ddot{S}_{\bf
q}^{+}| -\ddot{S}_{-\bf q}^{-}))_{\omega}^{(\rm proper)}\, ,
 \label{b4}
\end{equation}
where
\begin{equation}
(( A | B ))_{\omega} = - i \int_{0}^{\infty} dt  e^{i
   \omega t}  (A(t), B)
\label{b5}
\end{equation}
is the Kubo-Mori relaxation function and its  ``proper'' part
means that it does not contain parts connected by a single
relaxation function in the GMFA. The spin-fluctuation spectrum is
given by the imaginary part of the DSS  (\ref{b2}),
\begin{eqnarray}
\chi''({\bf q}, \omega)  =  \frac{- \omega \, \Sigma{''}({\bf
q},\omega)\; m({\bf q})} {[\omega^2 -  \omega_{\bf q}^2 - \omega
\, \Sigma{'}({\bf q},\omega)]^2  + [\omega \, \Sigma{''}({\bf
q},\omega)]^2} \, ,
 \label{b6}
\end{eqnarray}
where $\, \Sigma({\bf q},\omega +i0^+)= \Sigma{'}({\bf q},\omega)
+ i \Sigma{''}({\bf q},\omega)$,  and $\,\Sigma{'}({\bf
q},\omega) = - \Sigma{'}({\bf q}, -\omega)$ and $ \Sigma{''}({\bf
q},\omega) =  \Sigma{''}({\bf q}, - \omega) < 0$ are the real and
imaginary parts of the self-energy, respectively.

\subsection{Static properties}

To calculate the static susceptibility and  the spin-excitation
spectrum $\omega_{\bf q}$ in Eq.~(\ref {b3}), we  use the equality
\begin{equation}
m({\bf q}) = \langle[i\dot{S}_{\bf q}^{+},S_{-{\bf q}}^{-}]\rangle
=(-\ddot{S}_{\bf q}^{+},S_{-{\bf q}}^{-})  ,
 \label{b7}
\end{equation}
where
\begin{equation}
m({\bf q})= -8t(1-\gamma_{\bf q})F_{1,0}-8J(1-\gamma_{\bf
q})C_{1,0}
 \label{b8}
\end{equation}
with $\gamma_{\bf q}=(1/2)\,(\cos q_x + \cos q_y)$ (we take the
lattice spacing $a$ to be unity),
  $\, F_{n,m} \equiv F_{\bf R} = \langle X_{\bf 0}^{\sigma 0}\,
X_{\bf R}^{0\sigma}\rangle $,
 $\,  C_{n, m}  \equiv C_{\bf R} = \langle S^+_{\bf 0} \, S^-_{\bf R} \rangle$,
 and ${\bf R} = n e_x + m e_y \,$.
Here, we take into account the hopping integral $t_{ij}$ and the
exchange interaction $J_{ij}$ for the nearest neighbors only
denoted by $t$ and $J$, respectively.
\par
To calculate the correlation function $(-\ddot{S}_{\bf
q}^{+},S_{-{\bf q}}^{-})$ in Eq.~(\ref{b7}), we take the site
representation and use the decoupling procedure which is
equivalent to the mode-coupling approximation for the equal-time
correlation function (see Appendix~A). We obtain  $(-\ddot{S}_{\bf
q}^{+},S_{-{\bf q}}^{-}) = \omega_{\bf q}^2 \, ({S}_{\bf
q}^{+},S_{-{\bf q}}^{-})$ and, by comparison with Eq.~(\ref{b3}),
we get  the spin-excitation spectrum
\begin{eqnarray}
\omega_{\bf q}^2 &=& 8t^2\lambda_1(1-\gamma_{\bf
q})(1-n-F_{2,0}-2F_{1,1})
\nonumber \\
 &+ & 4J^2(1-\gamma_{\bf q})
(\lambda_2\frac{n}{2}-\alpha_1C_{1,0}(4\gamma_{\bf q}+1)
\nonumber\\
&+&\alpha_2(2C_{1,1}+C_{2,0})).
 \label{b9}
\end{eqnarray}
The decoupling parameters $\alpha_1, \alpha_2, \lambda_1$, and
$\lambda_2$ are explained in Appendix~A. Thus, the static
susceptibility can be calculated from Eq.~(\ref {b3}).
\par
The AF correlation length $\xi$ may be calculated  by expanding
the static susceptibility in the neighborhood of the AF wave
vector ${\bf Q} = (\pi,\pi)$ , $\chi_{\bf Q + \bf k}= \chi_{\bf
Q}/(1+\xi^2\, k^2)$.~\cite{Shimahara91,Winterfeldt97} We get
\begin{equation}
\xi^2=\frac{8J^2\alpha_1|C_{1,0}|}{\omega_{\bf Q}^2}.
 \label{b10}
\end{equation}
The critical behavior of the model  (\ref{b1}) is reflected by
the divergence of $\chi_{\bf Q}$ and $\xi$ as $T \rightarrow 0$,
i.e., by $\omega_{\bf Q}(T = 0) = 0$. In the phase with AF
long-range order (LRO) which, in two dimensions, may occur at $T
=0$ only, the correlation function $C_{\bf R}$ is written
as~\cite{Shimahara91,Winterfeldt97}
\begin{equation}
C_{\bf R} = \frac{1}{N}\sum_{{\bf q} \neq {\bf Q}}\,C_{\bf q}
 {\rm e}^{i {\bf q R}} + C\,{\rm e}^{i {\bf Q R}} ,
 \label{b11}
\end{equation}
where $C_{\bf q} = \langle S^+_{\bf q} S^-_{-\bf q}  \rangle $.
The condensation part $C$ determines the staggered magnetization
which is defined in the spin-rotation-invariant form
\begin{equation}
m^2 = \frac{3}{2N}\sum_{\bf R} \,C_{\bf R}
 {\rm e}^{- i {\bf Q R}} = \frac{3}{2} \, C.
 \label{b12}
\end{equation}
The static susceptibility, the correlation functions, the
correlation length, and  the magnetization  are calculated in the
GMFA for arbitrary   temperatures and doping (see Sec.~III.A).
Then, the GMFA results are used for the calculation of the
self-energy (see Sec.~III.B).

\subsection{Self-energy}

The self-energy  (\ref{b4}) can be written in terms of the
corresponding time-dependent correlation function as
\begin{eqnarray}
\Sigma({\bf q},\omega)= \frac{1}{2\pi m({\bf q})}
\int_{-\infty}^{\infty}d\omega^{\prime}\frac{e^{\beta\omega^{\prime}}-1}
{\omega^{\prime}(\omega-\omega^{\prime})}\nonumber\\
\,\int_{-\infty}^{\infty} dte^{i\omega^{\prime}t}\langle \,
\ddot{S}^{-}_{- \bf q}\; \ddot{S}^{+}_{\bf q}(t) \rangle^{\rm
proper}.
 \label{b13}
\end{eqnarray}
The self-energy is calculated in the mode-coupling approximation
for the multisite  correlation functions resulting from the
operator $\ddot{S}^{+}_{ \bf q}(t)$  as outlined  in Appendix~B.
We consider  only the imaginary part of the self-energy
(\ref{b13}) since the real part is given by the dispersion
relation~\cite{Zubarev60}.
\par
As it turns out by numerical evaluations  (see Sec.~III.B), the
largest contributions come from  two diagonal terms. For the
first term $\Sigma''_{J}$ we get
\begin{eqnarray}
&&\Sigma''_{J}({\bf q},\omega) = \frac{ \pi \,(2\,J)^4}{2 m({\bf
q})\,\omega \, N(\omega)}\, \frac{1}{N^2}\sum_{{\bf q}_1,{\bf
q}_2}\int_{-\infty}^{\infty} d \omega_1 d \omega_2
  \nonumber \\
&&\{\Gamma^2_{{\bf q}_1 {\bf q}_2 {\bf q}_3} + \Gamma_{{\bf q}_1
{\bf q}_2 {\bf q}_3}\Gamma_{{\bf q}_2 {\bf q}_1 {\bf q}_3}\}
 N(\omega_1 )N(\omega_2)
 \label{b16} \\
&& N(\omega-\omega_1-\omega_2)B_{{\bf q}_1}(\omega_1) B_{{\bf
q}_2}(\omega_2) B_{{\bf q}_3}(\omega- \omega_1 -\omega_2) ,
 \nonumber
\end{eqnarray}
where $N(\omega)=(e^{\beta\omega}-1)^{-1}$ and ${\bf q} = {\bf
q}_1 + {\bf q}_2 + {\bf q}_3\,$. The spectral density of the
spin-fluctuation spectrum $B_{\bf q}(\omega)=
(1/\pi)\,\chi''({\bf q}, \omega) $ is given by Eq.~(\ref{b6}). The
vertex for the spin-spin scattering reads (cf.
Ref.~\cite{Winterfeldt99})
\begin{eqnarray}
\Gamma_{{\bf q}_1 {\bf q}_2 {\bf q}_3}= 4 (\gamma_{{\bf q}_3+{\bf
q}_1} - \gamma_{{\bf q}_2}) (\gamma_{{\bf q}_3}-\gamma_{{\bf
q}_1})
\nonumber\\
-\gamma_{{\bf q}_1}+\gamma_{{\bf q}_3}+\gamma_{{\bf q}_2+{\bf
q}_3}-\gamma_{{\bf q}_2+{\bf q}_1}.
 \label{b18}
\end{eqnarray}
The  second term $\Sigma''_{t}$ is given by
\begin{eqnarray}
&&\Sigma''_{t}({\bf q},\omega)= \frac{\pi (2\, t)^4}{m({\bf q})\,
\omega \, N(\omega)}\, \frac{1}{N^2}
 \sum_{{\bf q}_1,{\bf q}_2}\,\int_{-\infty}^{\infty}d\omega_1 d\omega_2
\label{b19} \\
 &&\{ \Lambda^2_{{\bf q}_1 {\bf q}_2 {\bf q}_3}
+ \Lambda^2_{{\bf q}_3 {\bf q}_2 {\bf q}_1} \}
  N(\omega_2)n(\omega + \omega_1 - \omega_2)[1-
n(\omega_1)]
\nonumber\\
&& [(1/4) N_{{\bf q}_2}(\omega_2) + B_{{\bf q}_2}(\omega_2)]
A_{{\bf q}_1}(\omega_1) A_{{\bf q}_3}(\omega+\omega_1-\omega_2),
 \nonumber
\end{eqnarray}
where $n(\omega)= (e^{\beta\omega}+1)^{-1}$. Here, the
single-particle spectral function $\, A_{\bf q}(\omega)=
-(1/\pi){\rm Im} \langle\langle X^{0\sigma}_{\bf
q}|X^{\sigma0}_{\bf q}\rangle\rangle_{\omega}$ and the
charge-susceptibility spectral function $\,N_{\bf q}(\omega)=
-(1/\pi) {\rm Im} \langle\langle n_{\bf q}|n_{-{\bf
q}}\rangle\rangle_{\omega}$ are introduced. The vertex for the
spin-hole scattering reads
\begin{eqnarray}
 \Lambda_{{\bf q}_1 {\bf q}_2 {\bf q}_3}
= 4(\gamma_{{\bf q}_3+{\bf q}_2}
  -  \gamma_{{\bf q}_1 })\, \gamma_{q_3}
+ \gamma_{{\bf q}_2}-\gamma_{{\bf q}_1 + {\bf q}_3}.
 \label{b20}
\end{eqnarray}
The remaining terms in the self-energy  are considered in
Appendix~B. In Sec.~III.B we calculate the diagonal terms for
various doping and temperatures.
\par
We would like to emphasize that in our calculation of the
self-energy in Eq.~(\ref {b19}) contributions from the charge and
spin excitations are  taken into account explicitly by the
spectral densities $ N_{{\bf q}_2}(\omega_2)$ and $B_{{\bf
q}_2}(\omega_2)$. Contrary to this, in Refs.~\cite{Sega03} and
~\cite{Sherman03a} these terms   have been approximated by some
kind of static or mean-field-type expressions. This results in
the self-energy of the form similar to that given by a
conventional particle-hole loop diagram used in the weak coupling
theory like RPA. This form of the self-energy can be readily
obtained from our expression (\ref {b19}), if we disregard the
charge fluctuation contribution and neglect a small spin
excitation energy $\omega_2$ in comparison with the Fermi energy
in the Fermi function and in the hole spectral function: $\,
n(\omega + \omega_1 - \omega_2) A_{{\bf
q}_3}(\omega+\omega_1-\omega_2)
 \simeq n(\omega + \omega_1)
A_{{\bf q}_3}(\omega+\omega_1) \,$. Then we can integrate over
$\omega_2$ in Eq.~(\ref{b19}) which gives $\,
\int_{-\infty}^{\infty} d\omega_2   N(\omega_2) B_{{\bf
q}_2}(\omega_2) = C_{{\bf q}_2}, \,$ where $ C_{\bf q} = \langle
S^+_{\bf q} S^-_{-\bf q} \rangle$.  As a result the self-energy
takes the form
\begin{eqnarray}
&&\Sigma''_{t}({\bf q},\omega)= \frac{\pi (2\, t)^4}{m({\bf q})\,
\omega \, }\, \frac{1}{N^2} \sum_{{\bf q}_1,{\bf q}_2}\, C_{{\bf
q}_2}\{ \Lambda^2_{{\bf q}_1 {\bf q}_2 {\bf q}_3} +
\Lambda^2_{{\bf q}_3 {\bf q}_2 {\bf q}_1} \}
\nonumber \\
 &&\int_{-\infty}^{\infty}d\omega_1
[ n(\omega + \omega_1 )- n(\omega_1)]
 A_{{\bf q}_1}(\omega_1) A_{{\bf q}_3}(\omega+\omega_1),
 \label{b21}
\end{eqnarray}
which is similar to that found in the one-loop particle-hole
approximation used in  Refs.~\cite{Sherman03}-\cite{Sega03} and
~\cite{Sherman03a} and describes the damping due to the decay of
spin fluctuations into electron-hole excitations. The same result
can be deduced, if in the mode-coupling approximation  (see
Appendix, Eq.~(\ref {B2})) the time-dependent spin correlation
function is approximated by its static value: $\,\langle
X^{\sigma\sigma}_{{\bf k}_2}(t)\,
 X^{\sigma\sigma}_{-{\bf k}_2}\rangle \approx \langle
X^{\sigma\sigma}_{{\bf k}_2}\,X^{\sigma\sigma}_{-{\bf
k}_2}\rangle\,$. As our numerical calculations have shown, the
imaginary part of the  self-energy (\ref {b21}) is about twice as
large in comparison with  that given by Eq.~(\ref {b19}).

\section{Numerical results}

To investigate the magnetic properties of the $t$--$J$ model, in
particular the spin dynamics at arbitrary temperatures and hole
concentrations, we start from the GMFA. By Eq.~(\ref{b2}) we get
the DSS
\begin{equation}
\chi^{(0)}({\bf q}, \omega)= \frac{m({\bf q})}{2 \, \omega_{\bf
q}}
  \left( \frac{1} { \omega + \omega_{\bf q}} -
  \frac{1} { \omega - \omega_{\bf q} }  \right)
 \label{n1}
\end{equation}
with $m({\bf q})$  and $\omega_{\bf q} $  given by
Eqs.~(\ref{b8}) and ~(\ref{b9}), respectively, and the
correlation function
\begin{equation}
  C_{\bf q} = \langle S^+_{\bf q} S^-_{-\bf q}  \rangle =
\frac{m({\bf q})}{ 2 \, \omega_{\bf q}}\,\coth \frac{\beta \,
\omega_{\bf q}}{2} \,  .
 \label{n2}
\end{equation}
The electron Green function is calculated in the Hubbard~I
approximation which yields
\begin{equation}
 \langle \langle X_{\bf q}^{0\sigma } |
 X_{{\bf q}}^{\sigma0)}\rangle \rangle_{\omega} =
\frac{1- n/2}{\omega  - E_{\bf q}+ \mu }
 \label{n3}
\end{equation}
with $ E_{\bf q}=   - 4 (1- n/2)\, t\, \gamma_{\bf q}$. We get
\begin{equation}
F_{\bf q} = \langle X_{\bf q}^{\sigma 0} X_{{\bf q}}^{0\sigma}
\rangle  = (1- n/2)\, n(E_{\bf q}- \mu) \, .
 \label{n4}
\end{equation}
The chemical potential $\mu$ is calculated by the number
condition $n =(2/N) \sum_{\bf q}F_{\bf q} $.
\par
To go one step beyond the GMFA, we calculate the self-energy
[Eqs.~(\ref{b16})-(\ref{b20})] by inserting the GMFA results.
Moreover, for the spectral function $N_{\bf q}(\omega)$ of the
dynamic charge susceptibility appearing in Eq.~(\ref{b19}) we
take the GMFA result of Ref.~\cite{Jackeli99}:
\begin{eqnarray}
N_{\bf q}^{(0)}(\omega) &= & 8\, t\, F_{1,0} (1 - \gamma_{\bf q} )
\nonumber\\
 & \times &({1}/{\Omega_{\bf q} })
 [ \delta (\omega - \Omega_{\bf q})
 -  \delta (\omega + \Omega_{\bf q})],
\label{n4a}
\end{eqnarray}
where $\,  \Omega_{\bf q}^2 = 8 \, t^2 (1 - \gamma_{\bf q})
  (1 - n/2)\,$ in the leading order of doping.

\subsection{Static properties}

Considering first the static magnetic properties in the GMFA, we
have to solve numerically the coupled system of self-consistency
equations for the correlation functions $C_{\bf R} = (1/N)
\sum_{\bf q} C_{\bf q}\, {\rm e}^{i {\bf q R}}$ and for the
transfer amplitudes $F_{\bf R} = (1/N) \sum_{\bf q} F_{\bf q}\,
{\rm e}^{i {\bf q R}}$. In the long-range ordered phase,
Eq.~(\ref{b11}) and the additional equation $\omega_{\rm Q} = 0$
determining the condensation part $C$ must be taken into account.
To this end, the parameters $ \alpha_1, \, \alpha_2, \, \lambda_1
\,$, and $ \lambda_2\,$ have to be determined,  where the sum rule
\begin{equation}
  C_{0, 0} = \langle S^+_{0} S^-_{0}  \rangle =
  \frac{1}{2} \, (1 -\delta)
 \label{n5}
\end{equation}
must be fulfilled at arbitrary temperatures and hole doping.
\par
We fix the decoupling parameters as follows. The parameters $
\alpha_1$ and $ \alpha_2\,$ are determined in the Heisenberg
limit $( \delta =0)$ and their values are taken also at finite
$\delta $. For $ \delta =0$ we have $F_{\bf R \neq 0} = 0$ so
that the itinerant contribution to the spectrum (\ref{b9})
vanishes, and $\omega_{\bf q}$ agrees with the result of
Ref.~\cite{Winterfeldt97}, where we have to put $ \lambda_2 =1$,
as can be seen from Eq.~(\ref{A8}). At $T=0$ we fix $ \alpha_1$
and $ \alpha_2\,$ by the sum rule $ C_{0, 0} = 1/2$ and, as an
input, by the value of the nearest-neighbor correlation function
obtained by exact diagonalization (ED), $ C_{1,0}^{ED}( \delta
=0) = - 0.234$ (Ref.~\cite{Bonca89}). We get  $ \alpha_1 =2.285$
and $ \, \alpha_2= 2.548$. At finite temperatures we determine $
\alpha_1(T)$ and   $ \alpha_2(T)$ from the sum rule   and the
ansatz (cf. Refs.~\cite{Shimahara91,Winterfeldt97}) $r(T) \equiv
[ \alpha_1(T) - 1] / [ \alpha_2(T) - 1] = r(0) = 0.830$.
\par
Considering the parameters  $\lambda_1 \,$ and $ \lambda_2\,$ at
finite $\delta $, at $T = 0$ we fix them by the sum rule
(\ref{n5}) and by the ED value $ C_{1,0}^{\rm ED}( \delta
=0.0625; J/t = 0.4) = - 0.176$ (Ref.~\cite{Bonca89}). We get
 $ \lambda_1 =0.195$ and $ \, \lambda_2= 0.515$ with $\lambda_1 / \lambda_2 = 0.378$.
At arbitrary temperatures, doping, and ratios of $J/t$ we
determine $ \lambda_1 $ and $ \, \lambda_2$ from the sum rule
(\ref{n5}) and the ansatz
\begin{equation}
\lambda_1(T, \delta;\,J/t) / \lambda_2 (T, \delta;\,J/t)
 = 0.378 \, .
 \label{n6}
\end{equation}
\begin{figure}
\includegraphics[width=0.4\textwidth]{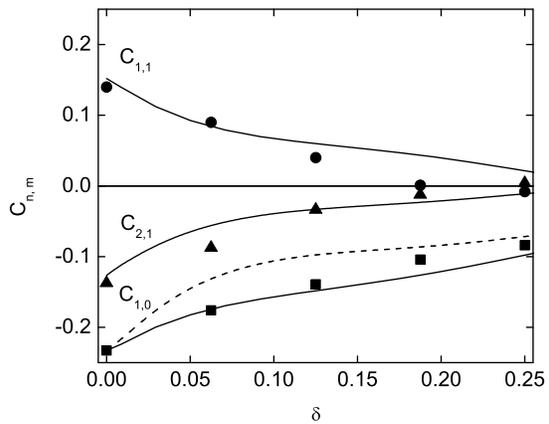}
\caption[]{ Spin correlation functions vs doping at $T = 0$ and
$J/t = 0.4$ (solid) and the ED data of Ref.~\cite{Bonca89}
(symbols). The  function $C_{1,0}$ at $J/t = 0.2$ is plotted by
the dashed line. }
 \label{fig1}
\end{figure}
In Fig.~\ref{fig1} our results for the doping dependence of the
spin correlation functions at $T = 0$ and $J/t = 0.4$ are
presented. They show a good agreement with the ED  data of
Ref.~\cite{Bonca89}. The different signs of $C_{n, m}$ reflect
the AF SRO  which gradually decreases with increasing doping and
decreasing ratios $J/t $.
\begin{figure}
\includegraphics[width=0.4\textwidth]{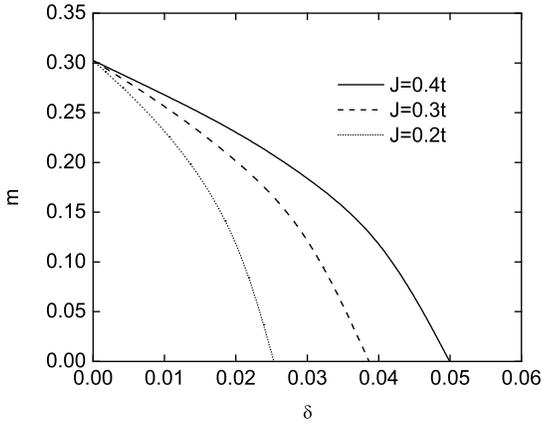}
\caption{ Staggered  magnetization as a function of doping for
different values of $J/t $.} \label{fig2}
\end{figure}
\par
Considering the staggered magnetization $m(\delta)$ at zero
temperature which is plotted in Fig.~\ref{fig2}, we obtain a
strong suppression of LRO   with increasing doping due to the
spin-hole interaction. In the Heisenberg limit we get $m(0)=
0.303$ which agrees with the value  $m(0)= 0.3074$ found in
quantum Monte Carlo (QMC)  simulations ~\cite{Wiese94}. At the
critical doping $ \delta_c(J/t)$ we obtain a transition from the
LRO phase to a paramagnetic phase with AF SRO. It is remarkable
that $ \delta_c$ is nearly proportional to $J/t$. This result
agrees with that found by the cumulant approach of
Ref.~\cite{Vojta96}, where our $ \delta_c$ values are somewhat
lower (e. g., in Ref.~\cite{Vojta96}, $ \delta_c \simeq 0.06$ for
$J/t = 0.4$). Note that the proportionality $ \delta_c \propto
J/t $ was not found in Ref.~\cite{Winterfeldt98}. The $ \delta_c$
values obtained are in qualitative agrement with neutron
scattering experiments on La$_{2-\delta}$Sr$_\delta$CuO$_4$
(LSCO) which reveal the vanishing of LRO at $ \delta_c \simeq
0.02$.~\cite{Kastner98}
\begin{figure}[ht!]
\includegraphics[width=0.4\textwidth]{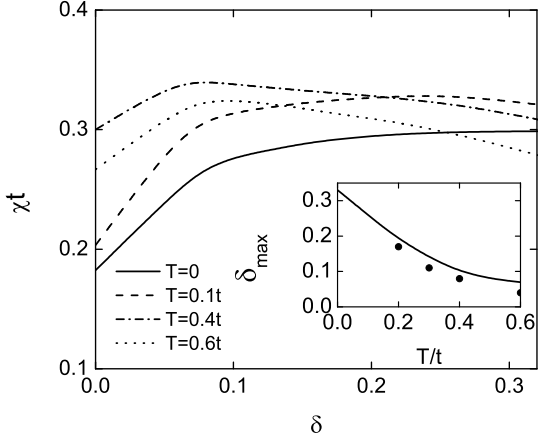}
\caption[]{ Uniform static spin susceptibility vs doping at $J/t
=0.3$ for various temperatures. The inset shows the position
$\delta_{\rm max}(T)$ of the maximum in $\chi$  vs $\delta$ in
comparison with the ED data (dots) of Ref.~\cite{Jaklic96}. \, .}
 \label{fig3}
\end{figure}
\par
In Fig.~\ref{fig3} the uniform static spin susceptibility $\chi =
(1/2)\lim_{q \rightarrow 0} \chi_{\bf q}$ at $J/t = 0.3$ is
plotted as a function of doping at various temperatures. Within
our theory, the increase of $\chi(\delta, T)$ upon doping is
caused by the decrease of SRO (cf. Fig.~\ref{fig1}),  i.e., of
the spin stiffness against orientation along a homogeneous
external magnetic field. At large enough doping,  $\chi$
decreases with increasing $\delta $ due to the decreasing number
of spins. The SRO-induced maximum of $\chi$ at $\delta_{\rm
max}(T)$ shifts to lower doping with increasing temperature, since
SRO effects are less pronounced at higher $T$. The doping
dependence of  $\chi$, especially the maximum at $\delta_{\rm
max}(T)$ (see the inset of Fig.~\ref{fig3}), is in accord with
the ED results of Ref.~\cite{Jaklic96}. Whereas the absolute
values of  $\chi$ turn out to be lower than the ED data, the
maximum position is in a remarkably good agreement with the ED
results. Note that in the spin-rotation invariant approach of
Ref.~\cite{Shimahara92} for $t \ll J$, the maximum  of  $\chi$ as
a function of doping was not reproduced. Our results are in
qualitative agreement with experiments on LSCO, where the
measured doping dependence of the magnetic susceptibility
exhibits a maximum at $\delta_{\rm max} \simeq 0.25$ over the
entire accessible temperature region, 50~K$\leq T
\leq$400~K.~\cite{Torrance89}
\par
Considering the temperature dependence of $\chi(T, \delta)$ at
fixed doping, from Fig.~\ref{fig3} it can be seen that there
appears a maximum at $T_{\rm max}(\delta)$ which shifts to lower
temperatures with increasing doping, in qualitative agreement
with the ED data~\cite{Jaklic96}. This maximum and the crossover
to the high-temperature Curie law $\chi \propto (1- \delta)/T$
can be understood as a SRO effect, in analogy to the explanation
of the doping dependence.
\par
\begin{figure}[ht!]
\includegraphics[width=0.4\textwidth]{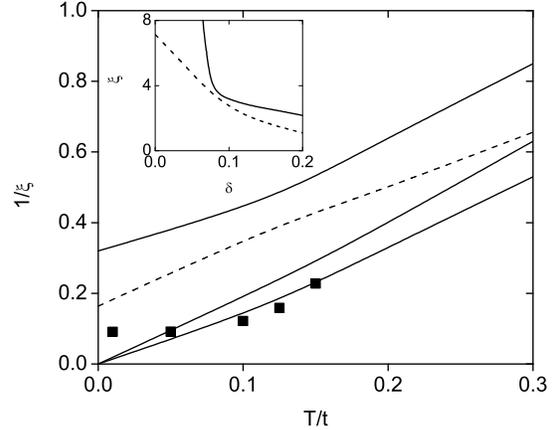}
\caption[]{Inverse AF correlation length vs $T$ at $J/t =0.4$
(solid lines) for doping $\delta = 0,\,0.04,\, 0.1 $, from bottom
to top,  and at  $J/t = 0.2 $ (dashed line) for $\delta= 0.04$.
The neutron-scattering data on
La$_{2-\delta}$Sr$_{\delta}$CuO$_{4}$ with $\delta = 0.04$ are
given by symbols \cite{Kastner98}. The inset exhibits the doping
dependence of the correlation length at $T=0$ (solid line) and
$T= 0.1t$ (dashed line) at $J/t =0.4$.}
 \label{fig4}
\end{figure}
Figure~\ref{fig4} shows the inverse correlation length
$\xi^{-1}(T, \delta, J/t)$. The qualitative behavior of $\,\xi
\,$ in the zero-temperature limit as function of doping and $J/t$
can be easily understood by considering the staggered
magnetization at $T = 0$ depicted in Fig.~\ref{fig2}. At a given
value of $J/t$ and $\delta < \delta_c$, in the limit $T \to 0$,
AF LRO emerges which is connected with the closing of the AF gap,
$\omega_{\bf Q} \to 0$, and, by Eq. (\ref{b10}), with the
divergence of $\,\xi \,$. At zero doping, $\xi^{-1}(T)$ exhibits
the known exponential decrease as $T \rightarrow
0$.~\cite{Shimahara91,Winterfeldt97} At $\delta > \delta_c$, the
ground state has no AF LRO, i.e., we have  $\omega_{\bf Q} > 0$,
and the correlation length saturates at  $T \to 0$. Equally,
taking $T = 0$, the transition from the AF LRO phase to a
paramagnetic phase with  AF SRO at $\delta = \delta_c$ is
accompanied by the change $\, \xi^{-1}(0, \delta < \delta_c, J/t)
=0\,$ to $\, \xi^{-1}(0, \delta > \delta_c, J/t)
> 0\,$. Considering the influence of the ratio $J/t$ on the
properties of $\xi$, let us compare the curves in Fig.~\ref{fig4}
for fixed $\delta = 0.04$ and $\, J/t = 0.4\,$ and $0.2$.
According to Fig.~\ref{fig2}, for $\, J/t = 0.4\,$ and $\, J/t =
0.2\,$ we have $\delta < \delta_c$  and $\delta > \delta_c$,  so
that $\, \xi^{-1}( 0, \delta, J/t = 0.4) =0\,$ and $\, \xi^{-1}(
0, \delta, J/t = 0.2) > 0\,$, respectively. The weakening of AF
correlations  (decrease of $\xi$) with decreasing exchange
interaction is in accord with the results for $C_{1.0}$  shown in
Fig.~\ref{fig1} for  $\, J/t = 0.4\,$ and $0.2$. Note that in
Ref.~\cite{Zavidonov98} a divergence of $\xi$ was found as $T
\rightarrow 0$ for arbitrary values of $\delta$, in disagreement
with experimental facts. This deficiency  may be due to employing
various decoupling schemes not used in  our theory.
\par
To compare the temperature dependence of $\xi^{-1}(T, \delta)$
with neutron-scattering experiments on LSCO at $T \leq
600$~K,~\cite{Kastner98} we take $a = 3.79$~\AA \ and $J =
130$~meV and consider the doping $\delta = 0.04$. As can be seen
in Fig.~\ref{fig4}, we obtain a reasonable  agreement with
experiments. Concerning the doping dependence of $\xi(\delta, T)$
depicted in the inset of Fig.~\ref{fig4}, it can be described
approximately by the proportionality $\xi(\delta, T) \propto
1/\sqrt{\delta}$ (dashed line)  which agrees with the
experimental findings.~\cite{Kastner98}

\subsection{Spin dynamics}

In this section we present results for the spin-fluctuation
spectra provided by the imaginary part of the DSS $\chi''({\bf
q}, \omega)$, Eq.~(\ref {b6}), where we neglect the real part of
the self-energy $\,\Sigma'({\bf q},\omega)\,$ (cf.
Ref.~\cite{Winterfeldt99}). The damping of spin fluctuations $\,
\Gamma({\bf q},\omega) $ is determined by the imaginary part of
the self-energy,  $\, \Gamma({\bf q},\omega)= -
(1/2)\Sigma''({\bf q},\omega)\,$, considered  in Sec.~II.C. Here
we mainly consider the damping at $\, \omega = \omega_{\bf q}, \;
\Gamma_{\bf q}= \Gamma({\bf q},\omega_{\bf q})  \,$. It turns out
that the major contributions to the damping are given by the
diagonal terms $\,\Sigma''_{J}({\bf q},\omega)\,$, Eq.~(\ref
{b16}), and $\,\Sigma''_{t}({\bf q},\omega)\,$, Eq.~(\ref {b19}),
while the interference terms, such  as $\,\Sigma_{Jt,Jt}''({\bf
q},\omega)\,$, Eq.~(\ref {B4}), appear to be much smaller and may
be neglected. That is, the damping $\,\Gamma_{\bf q} \,$ is the
sum of the spin-spin scattering contribution $\,\Gamma_{J, {\bf
q}} = - (1/2 )\Sigma''_{J}({\bf q},\omega_{\bf q}) \,$ and the
spin-hole scattering contribution $\,\Gamma_{t,{\bf q}} = - (1/2
)\Sigma''_{t}({\bf q},\omega_{\bf q}),\, \Gamma_{\bf q} =
\Gamma_{J,{\bf q}} +  \Gamma_{t,{\bf q}} \,$. Note that in
Refs.~\cite{Sega03} and ~\cite{Prelovsek04} the partition of the
damping into a spin-exchange contribution and a fermionic
contribution was suggested from the ED data. The numerical
calculations of  $\,\Sigma''({\bf q},\omega)\,$ are performed for
the exchange interaction $J = 0.3t$, the value which is usually
used in numerical simulations. This affords us to compare our
analytical results with finite cluster calculations and to check
the reliability of our approximations.
\par
\begin{figure}
\includegraphics[width=0.4\textwidth]{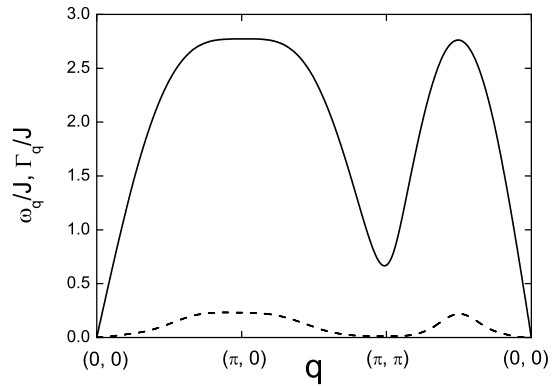}
\caption{Spectrum $\omega_{\bf q}$ (solid line) and damping
$\Gamma_{\bf q}$ (dashed line)
 in the Heisenberg limit, $\delta = 0$,  at $T=0.35 J$ .}
 \label{fig5}
\end{figure}
Let us first consider the Heisenberg limit $\delta = 0$.
Figure~\ref{fig5} shows the spin-excitation spectrum
$\omega_{\bf q}$, Eq.~(\ref {b9}),  and the damping $\, \Gamma_
{\bf q} = \Gamma_ {J, \bf q}$. The results are similar to those
obtained in Ref.~\cite{Winterfeldt99}. In the spin-wave region,
at $\,q \xi \gg 1$,  we get well-defined quasiparticles with
$\Gamma_ { \bf q} \ll \omega_{\bf q}$. For example, for ${\bf q}
= \pi\,(1/2, 1/2)$ and $T = 0.35 J$ we have $\, q \gg \xi^{-1}
\simeq 0.1$ (see Fig.~\ref{fig4}) and $\Gamma_{ \bf q}
/\omega_{\bf q} \simeq 0.1$. Well-defined spin excitations for
the two-dimensional Heisenberg model have been  found  by several
authors (for a review see Ref.~\cite{Manousakis91}). In
particular, as shown in Ref.~\cite{Tyc90}, if $T \to 0$  and $q
\to 0$ with the restriction  $q \xi \gg 1$, the ratio of the
damping to the spin-wave excitation energy  tends to zero:
$\Gamma_{J,{\bf q}}/ \omega_{{\bf q}} \rightarrow 0$.
\par
\begin{figure}
\includegraphics[width=0.4\textwidth]{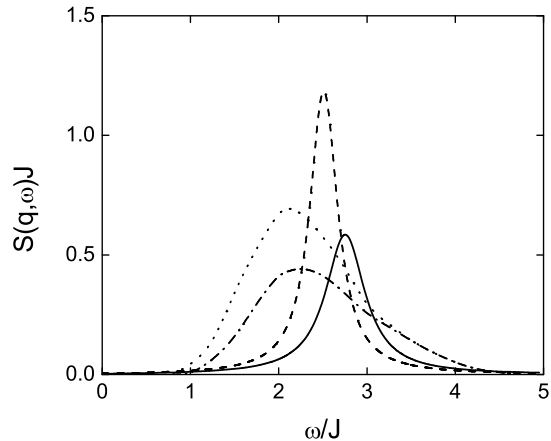}
\caption{Dynamic structure factor $S({\bf q}, \omega)$  in the
Heisenberg limit, $\delta = 0$,  at $T= 0.38J$ for  wave vectors:
${\bf q}_1 = \pi(1/2,1/2)$ (solid line) and  ${\bf q}_2  =
\pi(5/8, 5/8)$ (dashed line) in comparison with the QMC results
of Ref.~\cite{Makivic92} for ${\bf q}_1$ (dashed-dotted line) and
 for ${\bf q}_2 $ (dotted line).\,}
 \label{fig6}
\end{figure}
To compare our results for the damping with the QMC   data of
Ref.~\cite{Makivic92}, we have considered the linewidth
$\Lambda_{\bf q}$ of the relaxation function $\, F({\bf
q},\omega) = 4 [\beta\omega \chi_{ \bf q}]^{-1}\chi''({\bf
q},\omega) \,$  at $T = 0.35 J$, where $\Lambda_{\bf q} \simeq 2
\Gamma_{ \bf q} $ (Ref.~\cite{Winterfeldt99}), and have found a
good agreement. For a further comparison with the QMC data we
calculate the dynamic structure factor $\,S({\bf q}, \omega) = [1-
\exp(-\beta\omega]^{-1}\,\chi''({\bf q}, \omega)$. In
Fig.~\ref{fig6} our results  at $T = 0.38 J$ and for  two wave
vectors in the spin-wave region are plotted. The peaks in
$\,S({\bf q}, \omega) $, occurring nearly at $\omega_{\bf q}$
(cf.~Fig.~\ref{fig5}), reveal well-defined spin excitations.
Comparing the peak heights with the QMC values, we get a better
agreement than it was found in Ref.~\cite{Winterfeldt99}. This
may be ascribed to Eq.~(\ref {b16}) which corrects the result of
Ref.~\cite{Winterfeldt99} by the appearance of the additional
term $\,\Gamma_{{\bf q}_1 {\bf q}_2 {\bf q}_3}\Gamma_{{\bf q}_2
{\bf q}_1 {\bf q}_3}\,$ that cannot be written as a square of the
vertex $\Gamma_{{\bf q}_1 {\bf q}_2 {\bf q}_3}$.
\par
\begin{figure}[ht!]
\includegraphics[width=0.4\textwidth]{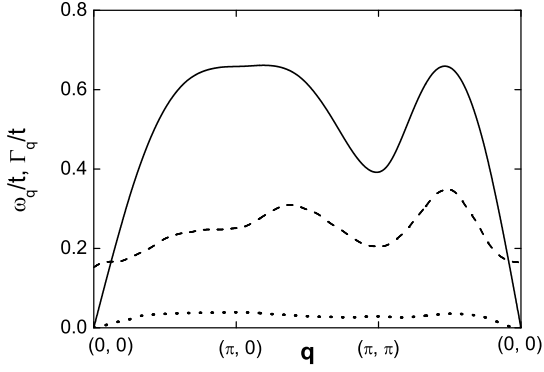}
\caption{Spectrum $\omega_{\bf q}$ (solid line), and  damping $\,
\Gamma_{J, \bf q}$ (dotted line) and   $\, \Gamma_{t, \bf q}\,$
(dashed line) at $T=0.15t$ and $\delta=0.1$.}
 \label{fig7}
\end{figure}
\begin{figure}[ht!]
\includegraphics[width=0.4\textwidth]{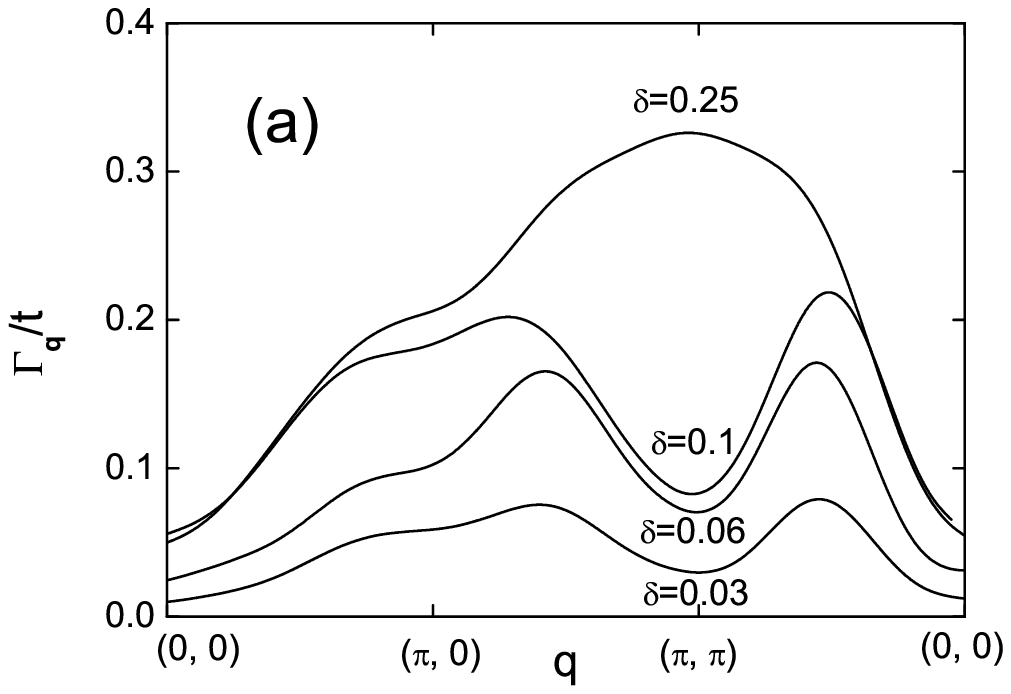}
\includegraphics[width=0.4\textwidth]{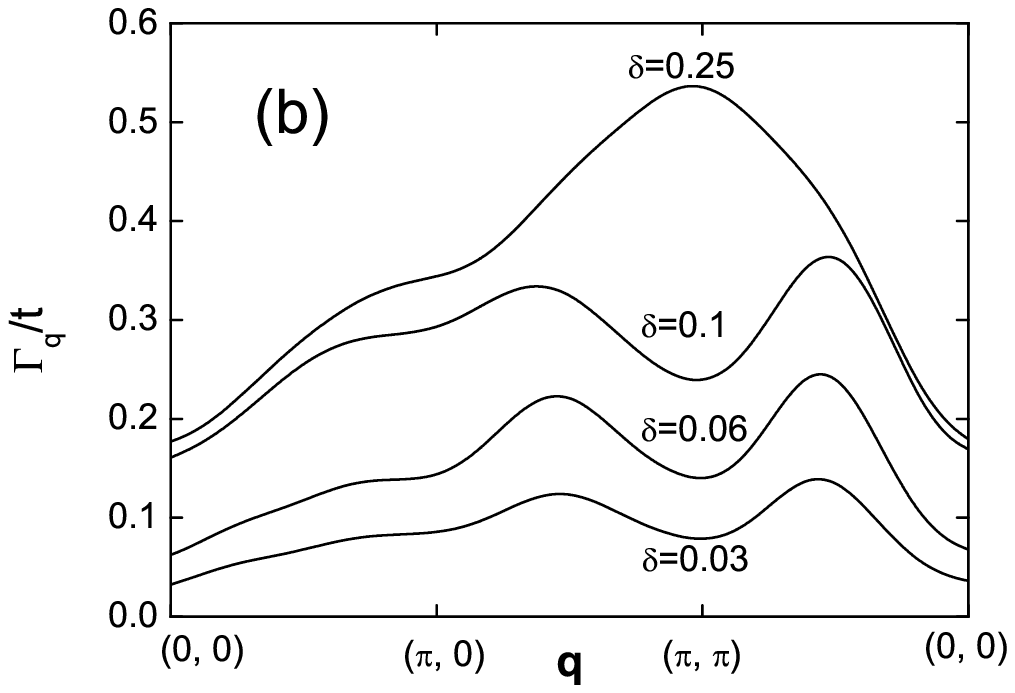}
\caption{Doping  dependence of damping $\Gamma_{\bf q}$ at
$T=0.1t$ (a) and $T=0.15t$ (b).}
 \label{fig8}
\end{figure}
For non-zero  doping the  spin-hole scattering contribution
$\,\Sigma''_{t}({\bf q},\omega)\,$, Eq.~(\ref {b19}), increases
rapidly with doping and temperature and already at moderate hole
concentration far exceeds the spin-spin scattering contribution
$\,\Sigma''_{J}({\bf q},\omega)\,$, Eq.~(\ref {b16}), as
demonstrated in  Figs.~\ref{fig7} and ~\ref{fig8}. Depending on
${\bf q}$, doping, and temperature, the spin excitations may have
a different character and dynamics. In particular, for the
spin-spin scattering contribution $\,\Sigma''_{J}({\bf
q},\omega)\,$ we observe, in the long-wavelength limit, $
\lim_{{\bf q} \to 0 } \Gamma_{J,{\bf q}} = 0$, as in the case of
the Heisenberg model shown in Fig.~\ref{fig5}.  Contrary to this
behavior, the damping $\Gamma_{t,{\bf q}}$, induced by the
spin-hole scattering, is finite in this limit, both taking
Eq.~(\ref{b19}) and Eq.~(\ref{b21}). The different behavior of
$\Gamma_{J, {\bf q}}$ and $\Gamma_{t, {\bf q}}$ may be explained
by the different ${\bf q}$-dependence of the spectral functions
entering Eqs.~(\ref{b16}) and (\ref{b19}). Whereas for spin
excitations the spectral function is proportional to $m({\bf
q})/\omega_{\bf q} \sim q $ for $q \to 0$ (see Eq.~(\ref {n1})),
for electrons it is finite in this limit (see Eq.~(\ref {n3})).
Therefore, in the limit of $\,{\bf q} = {\bf q}_1 + {\bf q}_2 +
{\bf q}_3 = 0\,$, for the spin-spin scattering the product
$m({\bf q}_1) m({\bf q}_2) m({\bf q}_3)/\omega_{{\bf
q}_1}\omega_{{\bf q}_2}\omega_{{\bf q}_3}$ gives a vanishingly
small contribution to the integrals over ${\bf q}_1,{\bf q}_2$ in
$\,\Sigma''_{J}({\bf q},\omega)\,$, Eq.~(\ref{b16}), while in the
spin-hole self-energy (\ref{b19}) there is no such a small
factor. The physical meaning of the finite damping $\Gamma_{t,
{\bf q}}$ at $q = 0$ can be explained similarly to that  of the
finite electrical conductivity in the low-frequency limit, which
is in fact a response function determining the damping of charge
fluctuations at ${\bf q}=0$. As  is well known, the relaxation
rate for the conductivity at zero frequency, i.e., the inverse
resistivity, is finite, if one takes into account momentum
relaxation of electron-hole pairs  on phonons.
\par
\begin{figure}[ht!]
\includegraphics[width=0.4\textwidth]{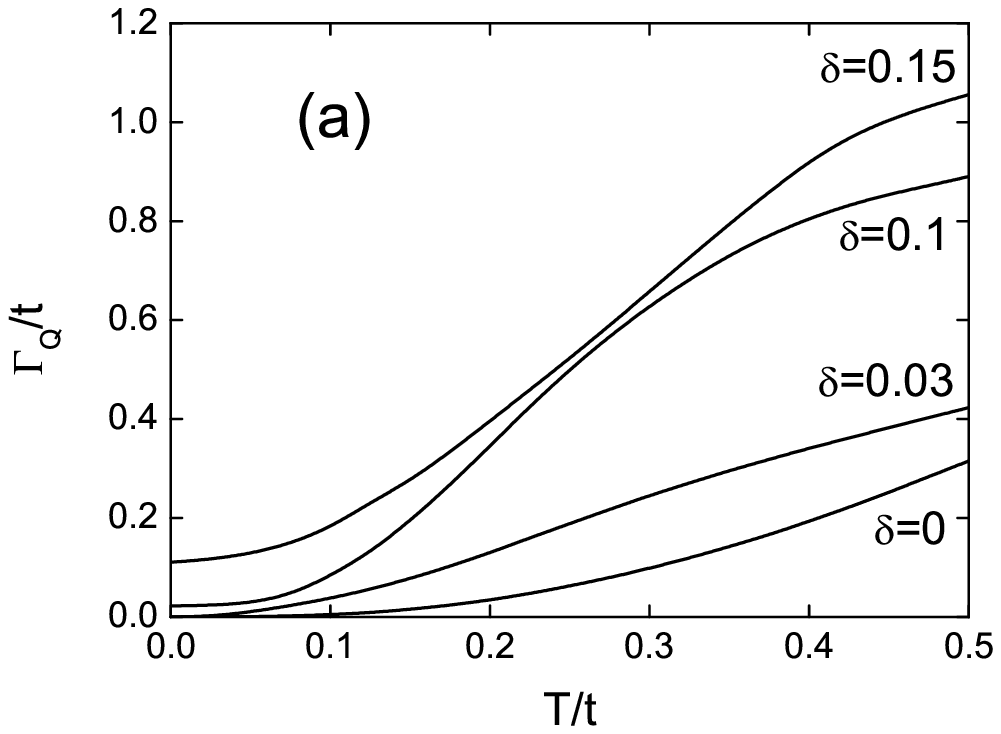}
\includegraphics[width=0.4\textwidth]{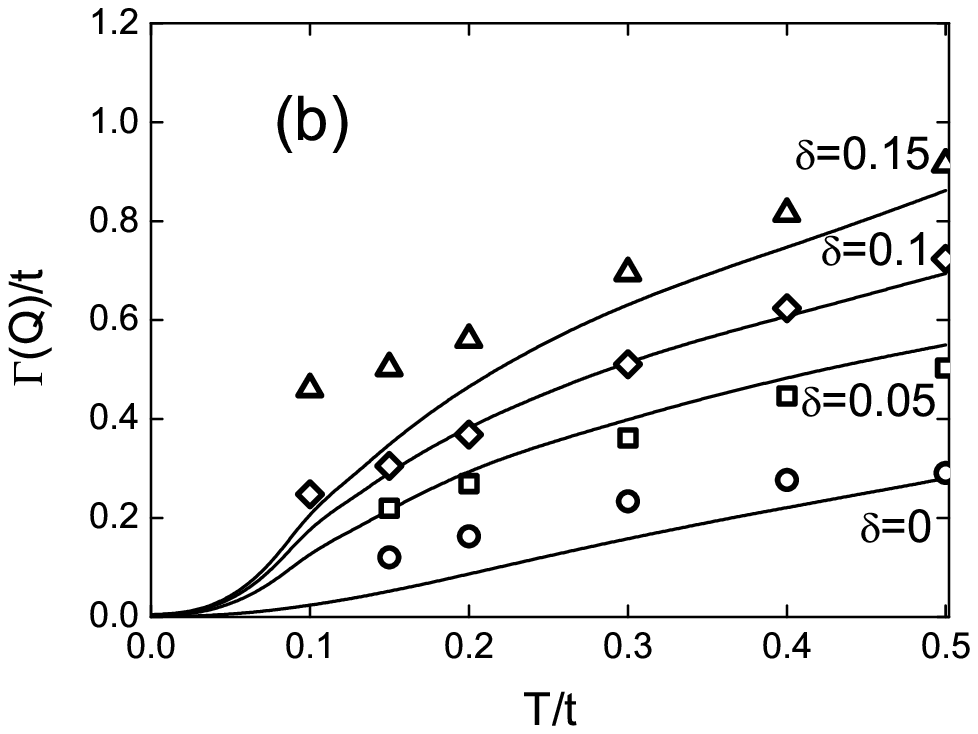}
\caption{Damping $\Gamma_{\bf Q} = - (1/2)\, \Sigma''({\bf Q},
\omega = \omega_{_{\bf Q}})$~(a) and low energy damping
$\Gamma({\bf Q}) = - (1/2)\, \Sigma''({\bf Q}, \omega = 0)$~(b)
as functions of temperature and doping in comparison with the ED
data of Ref.~\cite{Prelovsek04} (symbols). }
 \label{fig9ab}
\end{figure}
To discuss the temperature and doping dependence of the damping
(see Fig.~\ref{fig8}) in more detail, we choose ${\bf q = Q}\,$
and consider the damping $\Gamma_{\bf Q} = - (1/2)\,
\Sigma''({\bf Q}, \omega_{_{\bf Q}})$ as function of $T$ and
$\delta$  that is plotted in Fig.~\ref{fig9ab}~(a). In the
zero-temperature limit and for $ \delta < \delta_c $, where there
is AF LRO (see Fig.~\ref{fig2}), it can be shown analytically
that $\Gamma_{\bf Q}(T = 0, \delta < \delta_c ) = 0$.  That is,
in the LRO phase we get well-defined spin waves. The vanishing of
$\Gamma_{\bf Q}$ may be explained as follows. Spin excitations at
$T = 0$ can decay into particle-hole excitations with a positive
energy $\omega$ only to satisfy the energy conservation law.
Here, $\omega = \omega_{_{\bf Q}} = 0$ for  $\delta < \delta_c $
so that we have no damping. On the other hand, at $T = 0$ and
$\delta > \delta_c $, there is no LRO and $\omega_{_{\bf Q}} >
0$  which results in $\Gamma_{\bf Q}(T = 0,
 \delta > \delta_c ) > 0$ (see Fig.~\ref{fig9ab}~(a) for
$\delta = 0.1$ and 0.15). With increasing temperature and doping
the damping increases as expected. To compare our results with
the  data of Ref.~\cite{Prelovsek04}, in Fig.~\ref{fig9ab}~(b)
the temperature dependence of the low-energy damping  $\Gamma({\bf
Q}) = - (1/2)\, \Sigma''({\bf Q}, \omega = 0)$  for various
doping is shown. In Ref.~\cite{Prelovsek04} the function
$\,\gamma({\bf Q}) =  2 \Gamma({\bf Q})$ was extracted from the
finite-$T$ Lanczos data for the spectral function $\chi''(q,
\omega)$ using a simplified ansatz for the spin-excitation
spectrum, $\omega_{\bf q} \propto (({\bf q - Q})^2 +\kappa^2 )$
with $\kappa = \xi^{-1}$  taken as a temperature-independent
parameter, whereas our theory allows a direct microscopic
calculation of $\Gamma({\bf Q})$. In the high-temperature region
a remarkably good agreement is found. In the low-temperature
region, which is not accessible by the finite-$T$ Lanczos method,
the ED results were extrapolated to $T = 0$ with a finite value
of $\Gamma({\bf Q}; T = 0, \delta >  0)$. This is in contrast to
our result $\Gamma({\bf Q}; T = 0, \delta) = 0$ (cf.
Fig.~\ref{fig9ab}~(b)) which agrees with $\Gamma_{\bf Q}(T = 0,
\delta < \delta_c )= 0$ (cf. Fig.~\ref{fig9ab}~(a)), because both
quantities are calculated for $\omega = 0$, and which may be
understood as explained above.
\par
As illustrated in Fig.~\ref{fig8} and Fig.~\ref{fig9ab}, at low
enough doping and temperature, i.e., at small enough
$\,\Gamma_{t,{\bf q}}$, we may observe well-defined  high-energy
spin-wave-like  excitations with $\,q, k \gg 1/\xi \quad (k =
|{\bf q -Q }|)\,$ and $\,\Gamma_{{\bf q}} \ll \omega_{{\bf q}}\,$
propagating in AF SRO. Considering, for example, spin excitations
with ${\bf q} =(\pi, 0)$ at $\delta=0.1$ and $T = 0.15 t$, we
have $\,q \, \xi = 5.8$ with $ \, \xi = 1.85$ taken from
Fig.~\ref{fig4}, $\, \omega_{{\bf q}}= 0.66 t \,$, and
$\,\Gamma_{{\bf q}} = 0.27 t$ (see Figs.~\ref{fig7} and
\ref{fig8} (b)). That is, in this case we have strongly damped
spin waves.
\par
\begin{figure}[ht!]
\includegraphics[width=0.4\textwidth]{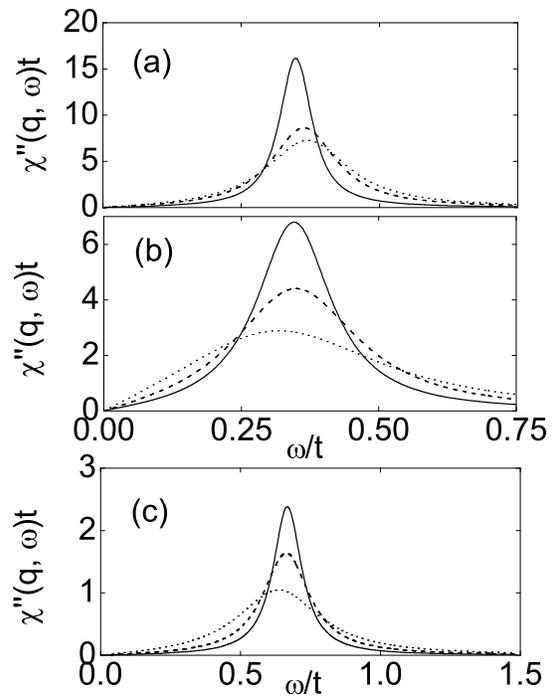}
\caption{Spectral function $\,\chi''({\bf q},\omega)\,$ for ${\bf
 q = Q} =(\pi, \pi)$  at $T = 0.1 t$ (a) and $T = 0.15 t$ (b) for
$\delta=0.03$ (solid line), $\delta=0.06$ (dashed line), and
$\delta=0.1$ (dotted line), and for ${\bf q} =(\pi, 0)$ at $T =
0.1 t$ (c) [note the change in the energy scale].}
 \label{fig9}
\end{figure}
\begin{figure}[ht!]
\includegraphics[width=0.4\textwidth]{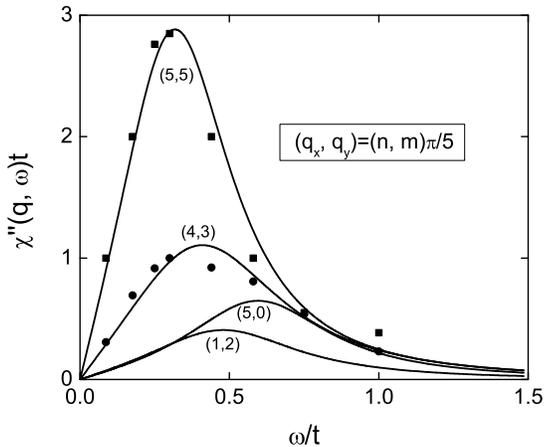}
\caption{Spectral function $\chi''({\bf q}, \omega)$ for various
wave vectors at $T=0.15t$ and $\delta=0.1$ in comparison with  ED
data (filled symbols, Ref.~\cite{Prelovsek04}).}
 \label{fig10}
\end{figure}

To discuss quantitatively the spectral function  $\,\chi''({\bf
q},\omega)\,$ shown in Figs.~\ref{fig9} and \ref{fig10}, in
particular the position of its maximum at $\omega_m$, we first
simplify Eq.~(\ref{b6}). By numerical evaluations, we have found
that the imaginary part of the self-energy only weakly depends on
frequency, which qualitatively agrees with the  results of
Ref.~\cite{Prelovsek04}. Therefore, we  put $\,- \Sigma''({\bf
q},\omega) = \eta_{\bf q} \simeq 2 \Gamma_{\bf q}$. Then, by
Eq.~(\ref{b6})  we get the resonance form
\begin{equation}
\chi''({\bf q}, \omega)  =  m({\bf q}) \frac{\eta_{\bf q} \omega}
 {(\omega^2 -  \omega_{\bf q}^2)^2  + \eta^2_{\bf q} \omega^2} \, ,
 \label{sd1}
\end{equation}
which has a maximum at $\omega^R_m$ given by
\begin{equation}
 \omega^R_m = \frac{1}{\sqrt{6}}\{2 \omega_{\bf q}^2 -
 \eta^2_{\bf q}+  [12 \omega_{\bf q}^4 + (2 \omega_{\bf q}^2 -
  \eta^2_{\bf q})^2]^{1/2}  \}^{1/2},
 \label{sd2}
\end{equation}
where $\lim_{\eta \to 0}\omega^R_m = \omega_{\bf q} $.
\par
Let us consider the region of low-frequency overdamped
spin-fluctuation modes playing an important role in the normal
phase of the cuprate superconductors, i.e., $\, \omega <
\omega_{\bf q} \ll \eta_{\bf q}\,$. Expanding Eq.~(\ref{sd1})
with respect to $\, \omega /\eta_{\bf q} < \omega_{\bf q}/
\eta_{\bf q} \ll 1\,$ and using Eq.~(\ref{b3}) we get
\begin{equation}
\chi''({\bf q}, \omega)  =  \chi_{\bf q}\,
\widetilde{\Gamma}_{\bf q} \,\frac{\omega}
 {\omega^2  + \widetilde{\Gamma}^2_{\bf q}} \, ;
 \quad \widetilde{\Gamma}_{\bf q}= \frac{\omega_{\bf q}^2}{\eta_{\bf
 q}},
 \label{sd3}
\end{equation}
where $\widetilde{\Gamma}_{\bf q} $ is the spin-fluctuation
excitation energy. Contrary to Ref.~\cite{Prelovsek04}, where a
similar expression was derived, we do not use an ansatz for the
spin-excitation spectrum  $\omega_{\bf q}$ (see above), but
calculate it microscopically by the GMFA [see Eq.~(\ref {b9})].
The Lorentzian (\ref{sd3}) has a maximum at $\omega^S_m =
 \widetilde{\Gamma}_{\bf q} $, which may be also obtained from the
 expansion of Eq.~(\ref{sd2}) with respect to $(\omega_{\bf q}/
\eta_{\bf q})^2$. The overdamped form corresponds to the
susceptibility $\, \chi({\bf q}, \omega) = \chi_{\bf q}\,
\widetilde{\Gamma}_{\bf q} ( \widetilde{\Gamma}_{\bf q} - i
\omega)^{-1}\,$ and, as a phenomenological ansatz, has been
frequently invoked in the study of cuprates, e.g., in the
calculation of the normal-state spin-fluctuation
conductivity.~\cite{Ihle94}
\par
To exemplify the spin dynamics in different regions, we first
consider spin waves with  ${\bf q} =(\pi, 0)$ at $\delta=0.1$ and
$T = 0.15 t$ (see above). We get (see  Fig.~\ref{fig10})
$\omega_m = 0.59 t \simeq \omega^R_m = 0.60 t\,$ and $ \omega^S_m
= 0.81 t\,$. That is, we find those excitations to have a
resonance character. On the other hand, the ED data of
Ref.~\cite{Prelovsek04} yield evidence for an overdoped spin
dynamics. This difference may be due to a slight underestimation
of the damping in our theory, which can be also seen in the more
pronounced peaks of the dynamic structure  factor at $\delta=0$
(see Fig.~\ref{fig6}) as compared with the QMC data. As seen in
Fig.~\ref{fig9}~(c), with increasing $\delta$ the maximum in
$\chi''({\bf q}, \omega) $ with  ${\bf q} =(\pi, 0)$ is shifted
to lower frequencies, in qualitative agreement  with the theory
of Ref.~\cite{Sherman03}.
\par
Next we consider the spectral function at ${\bf q = Q}$. At very
low doping, e.g., $\delta=0.03 $, and low enough temperature  the
damping $\Gamma_{\bf Q}$ is very small (see Fig.~\ref{fig8} and
Fig.~\ref{fig9ab}~(a)), where $\Gamma_{\bf Q} \ll \omega_{\bf
Q}$. In this case we observe underdamped spin modes characterized
by sharp resonance peaks in $\chi''({\bf Q}, \omega) $, as seen
in Fig\.~\ref{fig9}~(a),~(b). With increasing doping those modes
evolve into overdamped (relaxation-type) spin-fluctuation modes
(AF paramagnons) described by the broad spectrum (\ref{sd3}). For
example,  considering the AF mode at $\delta=0.1$ and $T=0.15t $
(see Figs.~\ref{fig7}, \ref{fig8}~(b), \ref{fig9}~(b), and
\ref{fig10}), we have $\omega_{\bf Q} = 0.4 t$, $\Gamma_{\bf Q} =
0.24 t $, and $ \omega_{m} = 0.32 t \simeq \omega_{m}^S = 0.33
t$. That is, the spectrum of this mode may be well described by
the overdamped form (\ref{sd3}).  As seen in Fig.~\ref{fig10},
our spin-fluctuation spectrum is in a remarkably good agreement
with the ED data of Ref.~\cite{Prelovsek04}. Let us consider the
shift of the maximum  in $\chi''({\bf Q}, \omega) $ at
$\omega_{m}$ with increasing doping at fixed temperature. As can
be seen from Fig.~\ref{fig9}~(a),~(b), at low (high)
temperatures, $\omega_{m}$ slightly increases (decreases)  with
doping, which results from the doping dependence of $\omega_{\bf
q}$ and $\Gamma_{\bf q}$. The increase of  $\omega_{m}$ with
$\delta$ at low $T$ is in qualitative agreement with the findings
of Ref.~\cite{Sherman03} ($T = 0.02 t$) and with experiments.
\par
\begin{figure}[ht!]
\includegraphics[width=0.4\textwidth]{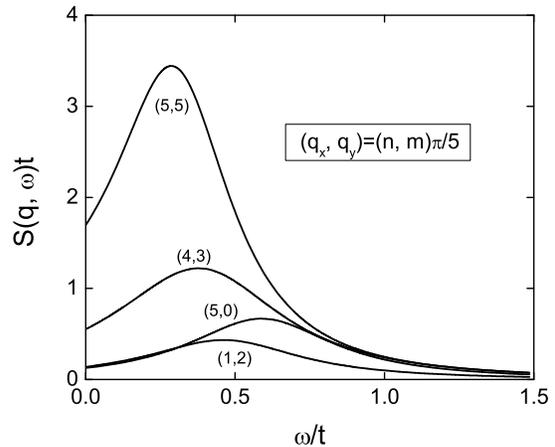}
\caption[]{Dynamic structure factor $S({\bf q}, \omega)$  at
$T=0.15t$ and $\delta=0.1$  for various wave vectors.}
 \label{fig11}
\end{figure}
In Fig.~\ref{fig11} the dynamic structure factor  $\,S({\bf q},
\omega)$, resulting from the spectral function shown in
Fig.~\ref{fig10}, is plotted. At $\omega =0$, by Eq.~(\ref {b6}),
we have $\,S({\bf q}, 0) = T \chi_{\bf q}\, \Sigma''({\bf q},0)
/\omega_{\bf q}^2 $, and for overdamped modes [Eq.~(\ref{sd3})]
we get $\,S({\bf q}, 0) = T \chi_{\bf q}/\widetilde{\Gamma}_{\bf
q} $. The shape of $\,S({\bf q}, \omega)$ for paramagnons  is in
a marked contrast to that for spin waves (compare also with
Fig.~\ref{fig6}).
\par
\begin{figure}[ht!]
\includegraphics[width=0.42\textwidth]{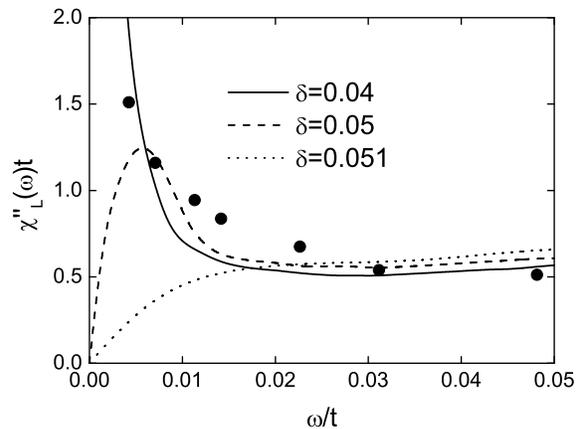}
\caption{Local spin susceptibility $\chi_L''(\omega)$  at $T = 0$
and small doping. The intensity  scaled experimental data on
La$_{1.96}$Sr$_{0.04}$CuO$_{4}$ at $T = 10$~K are shown by
dots~\cite{Keimer92}, where $t = 420$~meV is taken.}
 \label{fig13}
\end{figure}
Finally,  we present  results for the local susceptibility
\begin{equation}
\chi_L''(\omega) = \frac{1}{N}\, \sum_{\bf q} \chi''({\bf q},
\omega),
 \label{30}
\end{equation}
by using the data for $ \chi''({\bf q}, \omega)$,  Eq.~(\ref
{b6}). In Fig.~\ref{fig13} the local susceptibility at $T = 0 $
and small doping, $\delta = 0.04 - 0.051$, is shown. In the
neighborhood of the AF phase transition at $T = 0$ and $\delta =
\delta_c = 0.037$ (see Fig.~\ref{fig2}; $\omega_{\bf Q} = 0$),
the spin excitations with ${\bf q \simeq Q}$ are weakly damped
(see Fig.~\ref{fig9ab}~(a)). Therefore, at sufficiently low
$\delta $, the local susceptibility, which is just the density of
states (DOS) for spin-fluctuations, reveals a resonance maximum
at a frequency being close to $\omega_{\bf Q}$ with a high DOS
(see Fig.~\ref{fig13} at $\delta = 0.05$). Because $\omega_{\bf
Q}$ and $\Gamma_{\bf Q}$ decrease with decreasing  $\delta $, the
maximum shifts to lower frequencies and becomes very sharp (in
Fig.~\ref{fig13} at $\delta = 0.04$, only the upturn with
decreasing frequency is seen). On the other hand, with increasing
doping the damping becomes large enough to wash away the maximum
(cf. Fig.~\ref{fig13}). Note that we obtain, in addition to the
low-energy maximum, a broad maximum in $\chi_L''(\omega)$  at the
maximum energy of spin excitations, $\omega \sim 2J = 0.6 t$ (cf.
Fig.~\ref{fig5}). This feature was not found in
Ref.~\cite{Prelovsek04}, since a simplified  spin-excitation
spectrum $\omega_{\bf q}$ was used.
\par
\begin{figure}[ht!]
\includegraphics[width=0.4\textwidth]{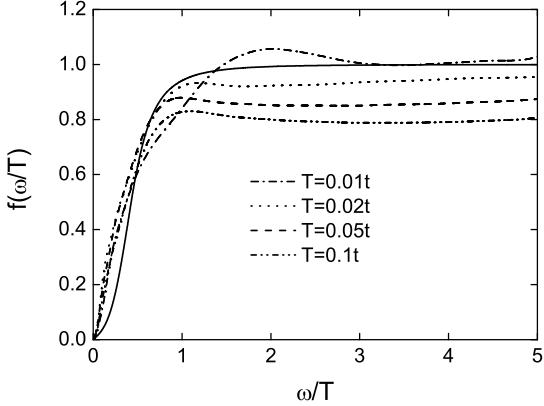}
\caption{Scaling function $f(\omega/T)$ for various temperatures
at doping $\delta = 0.04$. The solid line is the scaling function
found in the neutron-scattering experiments on
La$_{1.96}$Sr$_{0.04}$CuO$_{4}$ (Ref.~\cite{Keimer92}) and given
by Eq.~(\ref{31}).}
 \label{fig14}
\end{figure}
The pronounced upturn  behavior is observed in neutron-scattering
experiments on lightly doped cuprate compounds at low energies
and temperatures (see, e.g., Refs.~\cite{Keimer92} and
\cite{Stock08}). We get  a reasonable agreement  with experimental
data for La$_{1.96}$Sr$_{0.04}$CuO$_{4}$ ($\omega <
50$~meV)~\cite{Keimer92}, if we take the energy scale $t =
420$~meV (see Fig.~\ref{fig13}), which is the standard value of
$t \simeq 400$~meV in the $t$-$J$ model for cuprates. A much
better agreement with the data in Ref.~\cite{Keimer92} can be
obtained at the finite  temperature $T = 0.01 t$ but for the
energy scale $t = 1.1$~eV.  A qualitatively similar behavior has
been found in Ref.~\cite{Prelovsek04} within a
semi-phenomenological theory, where the agreement with experiment
was achieved by the choice $t \sim 0.1$~eV.
\par
Figure~\ref{fig14} shows the scaling function $f(\omega/T) =
\chi_L''(\omega,T)/\chi_L''(\omega,T=0) $.  The scaling behavior
is in a remarkable agreement with the data of the
neutron-scattering experiments on La$_{1.96}$Sr$_{0.04}$CuO$_{4}$
(Ref.~\cite{Keimer92}) which is shown by the solid line and
described by the function
\begin{equation}
 f\left(\frac{\omega}{T}\right) = \frac{2}{\pi}\,\arctan
\left[a_1\left( \frac{\omega}{T}\right) + a_2
\left(\frac{\omega}{T}\right)^3 \right],
 \label{31}
\end{equation}
with $a_1= 0.43$  and $a_2 = 10.5$. A similar scaling was
observed in the underdoped YBa$_2$Cu$_3$O$_{6.35}$ with the
parameters $a_1= 0.9$ and $a_2 = 2.8$ (Ref.~\cite{Stock08}). Our
results may be well approximated by the scaling function
(\ref{31})  with $a_1 =3$, but without the $({\omega}/{T})^3$
term, $a_2 = 0$,  that gives a nonlinear behavior at low values
of $({\omega}/{T}) \ll 1$. The weak non-monotonous behavior of
our scaling function at $\,T = 0.01 t \,$ results from the
appearance of a flat maximum in $\,\chi_L''(\omega, T)$ (see
above). Note that in Ref.~\cite{Prelovsek04}  the same scaling
function (\ref{31}) was found with $\, a_1 = 1.2, \,a_2 = 0 \,$,
that results in the saturation $f(\omega/T) \to 1$ at higher
values of $({\omega}/{T}) \gtrsim 2$ and in strong deviations
from the experiments on LSCO reported in Ref.~\cite{Keimer92},
but yields a good fit to experiments on Zn-substituted
YBa$_2$Cu$_3$O$_{6.6}$~\cite{Kakurai93}. This variation of the
scaling function may be explained by different values of doping
and of the corresponding parameters  determining the scaling
behavior (e.g., the AF correlation length $\xi$,
Ref.~\cite{Prelovsek04}).
\par
To sum up,  our studies of the DSS show a crossover  from
well-defined spin-wave-like excitations at low doping and
temperatures to  relaxation-type spin-fluctuation excitations
with increasing hole doping, which  is in agreement with
inelastic neutron-scattering experiments and numerical
simulations for  finite clusters. Moreover, we observe a
remarkable agreement of the scaling function with the data of
neutron-scattering experiments on LSCO.~\cite{Keimer92}

\section{Conclusion}

The relaxation-function  theory for the DSS in the $t$--$J$ model
in terms of Hubbard operators is formulated. By using a
spin-rotation-invariant theory for the DSS derived by us in
Ref.~\cite{Vladimirov05}, we calculate the static properties in
the GMFA similarly to Ref.~\cite{Winterfeldt98} and the
spin-fluctuation spectra. The mode-coupling approximation for the
force-force correlation functions, which take into account both
the exchange and kinetic contributions, was used. For the undoped
case described by the Heisenberg model our results are similar to
those in Ref.~\cite{Winterfeldt99} and for  finite doping they
show a reasonable agreement with available ED data and neutron
scattering experiments.
\par
Contrary to the previous studies based on the memory-function
method in Refs.~\cite{Sherman03} -- \cite{Prelovsek04}, we have
taken into account all contributions to the spin-excitation
spectrum $\omega_{\bf q}$ in the GMFA and to the self-energy
$\Sigma({\bf q},\omega)$ and thoroughly analyzed their
temperature and doping dependence. In particular, we have found
that the contribution from the hole-hopping term $\propto t^2$ in
the spectrum $\omega_{\bf q}$,  Eq.~(\ref {b9}), is large even in
the underdoped region, $\delta \lesssim 0.1$, and results in a
rapidly increasing with doping gap $\omega_{\bf Q}$ at the AF
wave vector ${\bf Q}$. This increase is much larger than in the
calculations in Ref.~\cite{Sherman03}, where the $\,t^2\,$
contribution has not been considered. We have also shown that the
largest contribution to the  self-energy $\Sigma({\bf q},\omega)$
comes from the hole-hopping term $ \Sigma''_{t}({\bf q},\omega)
\propto t^4$, Eq.~(\ref {b19}), at finite doping (see
Fig.~\ref{fig7}). This is in accord with Ref.~\cite{Sega03}, but
contrary to the approximation in Refs.~\cite{Sherman03},
\cite{Sherman04}, and \cite{Sherman03a}, where only the mixed
contribution $ \Sigma''_{Jt,Jt}({\bf q},\omega) \propto J^2t^2$
has been taken into account. The latter approximation should
strongly underestimate the damping of spin excitations at finite
doping. In our calculations of $ \Sigma''_{t}({\bf q},\omega)$ we
have also considered explicitly a contribution from spin
excitations in the self-energy (see Eq.~(\ref {b19})), while in
Refs.~\cite{Sega03} and \cite{Sherman03a} this contribution has
been considered in some kind of static  or mean-field-type
approximations.
\par
A comparison of the DSS  derived within the memory-function
approach, Eq.~(\ref{b6}), with the RPA form  $\, \chi({\bf
q},\omega) = \chi_0({\bf q},\omega)/[1- g_{\bf q} \, \chi_0({\bf
q},\omega)] \,$ (see, e.g., Ref.~\cite{Manske04}) has shown that
the RPA expression provides  reasonable results at large doping,
while in the underdoped region the RPA formula fails to describe
spin-wave-like excitations.~\cite{Prelovsek06} Whereas the
damping of spin excitations in Eq.~(\ref{b6}) at low doping is
quite small, e.g. $\,\Gamma_{\bf q}  \sim 0.2 t $ at $\delta =
0.1$ (see Fig.~\ref{fig7}), within the RPA it is much larger,
$\Gamma \sim t$. This results in overdamped spin dynamics
described by Eq.~(\ref{sd3})  even in the underdoped region.
Thus, we conclude that  the  relaxation-function approach is a
reliable theory for studying the spin dynamics in a broad region
of doping  and temperatures.
\par
In this paper we have not performed  a fully self-consistent
calculation of the electronic and spin-fluctuation spectra by
evaluating the spin correlation functions (\ref{n2}) in the GMFA
and the electron correlation functions (\ref{n4}) in the
Hubbard~I approximation. As was shown in Ref.~\cite{Plakida99},
static AF spin correlations and  self-energy effects result in a
strong renormalization of the electronic spectra and should be
taken into account in a consistent theory. This generalization
will be considered in a subsequent publication. The theory will
be also formulated  for the superconducting state by introducing
matrix electronic GF with normal and anomalous components as
given in Ref.~\cite{Plakida99}.

%------------------------------------

\acknowledgments

\indent The authors thank Ilya Eremin for valuable discussions.
Partial financial support by the Heisenberg--Landau Program of
JINR is acknowledged. One of the authors (N.P.) is grateful to
the  MPIPKS, Dresden, for the hospitality during his stay at the
Institute, where a  part of the present work has been done.

\appendix

\section{Decoupling procedure}

To calculate the correlation function
$(-\ddot{S}_{i}^{+},S_{l}^{-})$ in Eq.~(\ref {b7}) we consider
the equation
\begin{eqnarray}
 -\ddot{S}_{i }^{+} =[[{S}_{i}^{+},\, (H_{t} + H_{J})]\, ,
  \, (H_{t} +H_{J} )] \equiv \sum_{\alpha} F_{i}^{\alpha},
     \label{A1}
\end{eqnarray}
where $H_{t}$ and  $H_{J}$ are the hopping  and the exchange
parts of the Hamiltonian~(\ref {b1}) and $\alpha = tt,\, tJ,\, Jt,
\, JJ$. Here we have
\begin{eqnarray}
F_{i}^{tt}&=
&\sum_{j,n}t_{ij}\Bigl\{t_{jn}\left[H^-_{ijn}+H^+_{nji}\right]
-  (i \Longleftrightarrow j ) \Bigr\},  \qquad \label{A2} \\
F_{i}^{JJ} & = &
\frac{1}{4}\sum_{j,n}J_{ij}\Bigl\{J_{jn}\left[2P_{ijn}+\Pi_{ijn}\right]
  - (i \Longleftrightarrow j )\Bigr\}, \qquad \label{A3} \\
 F_{i}^{Jt}& = & [[{S}_{i}^{+},\,  H_{J}]\, , \, H_{t} ],\qquad
 F_{i}^{tJ}= [[{S}_{i}^{+},\,  H_{t}]\, , \, H_{J} ],
 \label{A3a}
\end{eqnarray}
where
\begin{eqnarray}
H^{\sigma}_{ijn}&=&X_{i}^{\sigma 0}X_{j}^{+-}X_{n}^{0\sigma} +
X_{i}^{+0}(X_{j}^{00}+X_{j}^{\sigma\sigma})X_{n}^{0-}, \quad
 \label{A4}\\
P_{ijn}&=&S_{i}^{z}S_{j}^{z}S_{n}^{+}+ S_{n}^{+}S_{i}^{z}S_{j}^{z}
\nonumber\\
& - & S_{i}^{z}S_{j}^{+}S_{n}^{z}- S_{n}^{z}S_{i}^{z}S_{j}^{+},
\quad \label{A5}\\
\Pi_{ijn}&=&S_{i}^{+}S_{j}^{-}S_{n}^{+}+S_{n}^{+}S_{i}^{+}S_{j}^{-}
\nonumber\\
& -& S_{i}^{+}S_{j}^{+}S_{n}^{-}-S_{n}^{-}S_{i}^{+}S_{j}^{+} \, .
\quad \label{A6}
\end{eqnarray}
Explicit expressions for $F_{i}^{tJ},  F_{i}^{Jt} $ are given in
Ref.~\cite{Vladimirov05}.
\par
To evaluate the corresponding multiparticle  correlation
functions in $(-\ddot{S}_{\bf q}^{+},S_{-{\bf q}}^{-})$ we
perform the following  decoupling procedure similar to that
proposed in
Refs.~\cite{Shimahara91,Shimahara92,Winterfeldt98,Winterfeldt99}
preserving the local correlations. The correlation functions from
$H^{\sigma}_{i j n}$ are decoupled  as
\begin{equation}
( X_{i}^{\sigma 0}X_{j}^{+-}X_{n}^{0 \sigma},S_{l}^{-}) =
   \lambda_1 \,\langle X_{i}^{\sigma 0} X_{n}^{0 \sigma}\rangle \, (S_{j}^{+},S_{l}^{-}),
 \label{A7}
\end{equation}
where, for $\, n = i,\, $  $ \,\langle X_{i}^{\sigma 0} X_{i}^{0
\sigma}\rangle =\langle X_{i}^{\sigma \sigma}\rangle = n/2\,$ and
the second term of $\,H^{\sigma}_{ijn}\,$ with $ n \neq  i$  is
neglected (cf. Ref.~\cite{Winterfeldt98}). Decoupling the
correlation functions from $\Pi_{i j n= i}$ and $ P_{ij, n= i}$
we introduce the parameter $\lambda_2$:
\begin{eqnarray}
&&(\Pi_{i j i},S_{l}^{-}) = - ( [X_{i}^{++} + X_{i}^{-
-}]\,S_{j}^{+},S_{l}^{-})
 \nonumber\\
&& = - \lambda_2\,\langle[X_{i}^{++} + X_{i}^{- -}]\rangle\,
(S_{j}^{+},S_{l}^{-}) ;
 \nonumber\\
&&( P_{i j i},S_{l}^{-}) =
 - (1/2) \,([X_{i}^{++} + X_{i}^{--}] S_{j}^{+},S_{l}^{-})
\nonumber \\
 &&= - \lambda_2\, (1/2) \, \langle[X_{i}^{++} + X_{i}^{--}] \rangle \,
(S_{j}^{+},S_{l}^{-}),
 \label{A8}
\end{eqnarray}
where we used the equations: $ \, S_{i}^{+}S_{i}^{-} =
X_{i}^{++},\; S_{i}^{z}S_{i}^{+}  + S_{i}^{+} S_{i}^{z}  = 0, \,$
and $\, ( S_{i}^{z})^2 = (1/4)(X_{i}^{++} + X_{i}^{--})\, $. Here
we take $ \lambda_2 \neq \lambda_1\,$, in contrast to the
approach of Refs.~\cite{Winterfeldt98,Vladimirov07}, $ \lambda_2
= \lambda_1\,$. In the Heisenberg limit $\delta = 0$ we have $\,
X_{i}^{++} + X_{i}^{--} \equiv 1$ so that  $\, \lambda_2 = 1\,$.
The parameters $\lambda_1, \lambda_2$ describe the
renormalization  of the vertex for spin scattering on charge
fluctuations.
\par
Considering the correlation functions from $\Pi_{ij n \neq i}$
and $P_{i j n \neq i}$, where $\{i, j, n \}$ forms a
nearest-neighbor sequence, we apply the decoupling scheme used in
Refs.~\cite{Shimahara91,Winterfeldt98,Winterfeldt99}:
\begin{eqnarray}
(S_{i}^{+}S_{j}^{+}S_{n}^{-},S_{l}^{-})& =
 & \alpha_1\,\langle S_{j}^{+}S_{n}^{-}\rangle\,(S_{i}^{+},S_{l}^{-})
 \nonumber\\
&+& \alpha_2\, \langle S_{i}^{+}S_{n}^{-}\rangle\,
(S_{j}^{+},S_{l}^{-}).
 \label{A9}
\end{eqnarray}
Here,  the parameters $\alpha_1$ and $ \alpha_2$ are attached to
nearest-neighbor and further-distant correlation functions,
respectively, and describe the renormalization of the vertex for
spin-spin scattering.  The determination of all parameters is
considered in Sec.~III.A.
\par
Calculating the spin-excitation spectrum and the  static
susceptibility (\ref {b3}) with Eq.~(\ref {b7}) we may take into
account only the diagonal contributions (\ref {A2}) and (\ref
{A3}) and omit the mixed contributions corresponding to $\,
F_{i}^{Jt}\,$ and $\, F_{i}^{tJ}\,$ in Eq.~(\ref {A3a}) according
to the following reasoning. The mixed contribution of the type
$\, (F_{i}^{Jt},\, S_{l}^{-})\,$ in the GMFA is proportional to
the difference of the correlation functions of the form $\,
\langle X^{++}_i - X^{--}_i \rangle \,$ or $\,\langle X^{0-}_i
X^{0-}_j - X^{0+}_i X^{0+}_j \rangle\,$, which vanishes in the
paramagnetic phase. In the same approximation  the mixed
contribution of the type $\, (F_{i}^{tJ}, S_{l}^{-})\,$ turns
out  to be proportional to higher-order correlation functions of
the type $( X_{i}^{+ 0} X_{j}^{0-}, \, S_{l}^{-})\,$ and may be
neglected.

\section{Mode-coupling approximation}

Using Eq.~(\ref {A1})  the two-time correlation function in
Eq.~(\ref{b13}) yields 16 terms for the self-energy $\Sigma({\bf
q},\omega) = \sum_{\alpha, \beta} \Sigma_{\alpha, \beta}({\bf
q},\omega)$, where, e.g.,
 $\Sigma_{JJ,JJ}({\bf q},\omega) = (( F^{JJ}_{\bf q}|
( F^{JJ}_{\bf q})^{+} ))_{\omega}/m({\bf q})$. In the site
representation of $\ddot{S}_{\bf q}^{+}$ in Eq.~(\ref{b13}) given
by Eqs.~(\ref{A1})--(\ref{A6}) we take into account products of
three spin operators on different sites only. This is clear in
the Heisenberg limit~\cite{Winterfeldt99}, where terms with
coinciding sites reduce to single operators and the proper part
of the correlation function in $\Sigma_{JJ,JJ}$ is considered.
For finite doping, as revealed by numerical evaluations, the
exclusion of terms with coinciding sites yields a better
agreement with exact data than the inclusion of those terms. We
calculate the two-time correlation functions  in the
mode-coupling approximation (see, e.g.,
Ref.~\cite{Winterfeldt99}), i.e., we approximate them  by a
product of three single-particle two-time correlation functions
as follows:
\begin{eqnarray}
&&\langle S^z_{{\bf k}_1}(t)S^z_{{\bf k}_2}(t)S^-_{{\bf k}_3}(t)
\, S^z_{{\bf k}'_1} S^z_{{\bf k}'_2} S^+_{{\bf k}'_3}\rangle
  \nonumber \\
&=&\langle S^z_{{\bf k}_1}(t)\, S^z_{-{\bf k}_1}\rangle\langle
S^z_{{\bf k}_2}(t)\,S^z_{-{\bf k}_2}\rangle \langle S^-_{{\bf
k}_3}(t)S^+_{-{\bf k}_3}\rangle
\nonumber \\
 &\times& (\delta_{{\bf k}_1, -{\bf k}'_1 } \delta_{{\bf k}_2, -{\bf k}'_2 } +
\delta_{{\bf k}_1, -{\bf k}'_2 } \delta_{{\bf k}_2, -{\bf k}'_1
})\,
  \delta_{{\bf k}_3, -{\bf k}'_3 },
 \label{B1}
\end{eqnarray}
\begin{eqnarray}
&&\langle X^{0+}_{{\bf k}_1}(t)X^{\sigma\sigma}_{{\bf
k}_2}(t)X^{-0}_{{\bf k}_3}(t) \,  X^{+0}_{{\bf
k}'_1}X^{\sigma\sigma}_{{\bf k}'_2}X^{0-}_{{\bf k}'_3} \rangle
 \nonumber \\
& = &  \langle X^{0+}_{{\bf k}_1}(t)\,X^{+0}_{{\bf k}_1}\rangle
 \langle X^{\sigma\sigma}_{{\bf k}_2}(t)\,X^{\sigma\sigma}_{-{\bf k}_2}\rangle
 \langle X^{-0}_{{\bf k}_3}(t)\, X^{0-}_{{\bf k}_3}\rangle  \nonumber \\
&&  \qquad  \times \; \delta_{{\bf k}_1, {\bf k}'_1 }
\,\delta_{{\bf k}_2, -{\bf k}'_2 }
  \, \delta_{{\bf k}_3, {\bf k}'_3 } .
 \label{B2}
\end{eqnarray}
In Ref.~\cite{Vladimirov05} we have shown that in the  Born
approximation (i.e., in the second order of the effective
vertices $t^2, J^2, tJ$) only six contributions to the
self-energy may be retained, so that
\begin{eqnarray}
\Sigma({\bf q},\omega)&= &\Sigma_{JJ,JJ}({\bf q},\omega)
  + \Sigma_{tt,tt}({\bf q},\omega)
 +\Sigma_{tJ,tJ}({\bf q},\omega)
 \nonumber \\
 & +& \Sigma_{Jt,Jt}({\bf q},\omega)
 + 2 \, \Sigma_{tJ,Jt}({\bf q},\omega).
 \label{B3}
\end{eqnarray}
The imaginary parts of the diagonal terms $\Sigma_{JJ,JJ} \equiv
\Sigma_{J} $ and $ \Sigma_{tt,tt}\equiv  \Sigma_{t}$ are given by
Eqs.~(\ref{b16}) and (\ref{b19}). For one of the interference
terms we obtain
\begin{eqnarray}
&&\Sigma_{Jt,Jt}''({\bf q},\omega)= \frac{\pi (2\, t)^2
\,(2\,J)^2}
 {m({\bf q})\,\omega\,N(\omega)} \frac{1}{N^2}
 \sum_{{\bf q}_1, {\bf q}_2}  \Gamma_{{\bf q}_1 {\bf q}_2 {\bf q}_3}^2
\nonumber\\
&& \int_{-\infty}^{\infty}d\omega_1 d\omega_2
N(\omega_1)n(\omega_2)n(\omega-\omega_1-\omega_2)
\nonumber\\
&&B_{{\bf q}_1}(\omega_1)A_{{\bf q}_2}(\omega_2)A_{{\bf
q}_3}(\omega-\omega_1-\omega_2).
 \label{B4}
\end{eqnarray}
with $\Gamma_{{\bf q}_1 {\bf q}_2 {\bf q}_3}$ given by
Eq.~(\ref{b18}), where the contributions linear in $\gamma_{\bf
q}$ reflect the exclusion of terms in $\ddot{S}_{i }^{+}$ with
coinciding sites.


\begin{thebibliography}{99}
\bibitem{Chubukov02} A.V. Chubukov, D. Pines, and J. Schmalian, in:
 {\it The Physics of Conventional and Unconventional
Superconductors}, edited by K.H.~Bennemann and J.B.~Ketterson,
(Springer Verlag, Berlin, 2004) Vol.~I, p.~495.
(arXiv:cond-mat/0201140).
%``A Spin Fluctuation model for  $d$-wave Superconductivity''.
\bibitem{Kastner98}  M.A. Kastner, R.J. Birgeneau,  G. Shirane, and
Y. Endoh, Rev. Mod. Phys.   {\bf 70}, 897 (1998).
% Magnetic, transport and optical properties of monolayer copper oxides
\bibitem{Bourges98} P. Bourges, in: {\it The
Gap Symmetry and Fluctuations in High Temperature
Superconductors}, edited by J. Bok, G. Deutscher, D. Pavuna, and
S.A. Wolf (Plenum Press, 1998), p. 349.
\bibitem{Manousakis91} E. Manousakis, Rev. Mod. Phys. {\bf 63},  1 (1991).
% The spin-1/2 Heisenberg antiferromagnet on a square lattice and
% its application to the cuprous oxides.
\bibitem{Manske04} D. Manske,  I. Eremin, and K.H.~Bennemann,
in: {\it The Physics of Conventional and Unconventional
Superconductors}, edited by  K.H.~Bennemann and J.B.~Ketterson
(Springer-Verlag, Berlin, 2002) Vol.~II, p.~731.
%``Electronic Theory for Superconductivity in  High-$T_c$ Cuprates
% and Sr$_2$RuO$_4$''.
\bibitem{Hubbard65}  J. Hubbard, Proc. R. Soc. London, Ser. A
 {\bf 285}, 542 (1965).
\bibitem{Anderson87} P.W. Anderson,
Science {\bf 235}, 1196 (1987).
\bibitem{Izyumov97}  Yu. A. Izyumov, Usp. Fiz. Nauk {\bf 167}, 465
(1997) [Phys.- Usp. {\bf 40}, 445 (1997)].
\bibitem{Plakida06}   N.M. Plakida, Fiz. Nizk. Temp.
(Low Temp. Phys., Ukraine)  32,  483 (2006).
% Theory of antiferromagnetic pairing in cuprate superconductors
%(Review article).
\bibitem{Izyumov90}  Yu. A. Izyumov and B. M. Letfulov,
 J. Phys.: Condens. Matter,  {\bf 2}, 8905 (1990);
  Yu.~A.~Izyumov and J. A. Hedersen, Int. J. Mod.
Phys. B {\bf 8}, 1877 (1994).
\bibitem{Dagotto94}  E. Dagotto, Rev. Mod. Phys {\bf 66}, 763 (1994).
\bibitem{Jaklic00}  J. Jakli\v{c} and P. Prelov\v{s}ek,
Advances in Physics   {\bf 49}, 1 (2000).
% Finite-temperature properties of doped antiferromagnet
\bibitem{Eder95}  R. Eder, Y Ohta, and S. Maekawa,
Phys. Rev. Lett.   {\bf 74}, 5124 (1995).
%Anomalous Spin and charge dynamics of the t-j model at low doping
\bibitem{Zubarev60}  D.N. Zubarev,
 %Double-time Green's functions in statistical  physics,
 Sov. Phys. Uspekhi {\bf 3}, 320 (1960).
\bibitem{Plakida73}  N.M.Plakida,  Phys. Lett. A  {\bf 43}, 481
(1973). % Dyson equation for Heisenberg ferromagnet.
\bibitem{Tserkovnikov81} Yu.A. Tserkovnikov,  Theor. Math. Phys.
{\bf 49}, 993 (1981);  Theor. Math. Phys. {\bf 52}, 712 (1982).
 %Teor. Mat. Fiz. {\bf 49}, 993 (1981);
 %Teor. Mat. Fiz. {\bf 52}, 147 (1982)
\bibitem{Jackeli98}  G. Jackeli and  N. M. Plakida,
Theor. Math. Phys. {\bf 114}, 335 (1998).
\bibitem{Sherman03}  A. Sherman and M. Schreiber,
Eur. Phys. J. B {\bf 32}, 203 (2003);
% ( Two-dimensional t-J model at moderate doping).
\bibitem{Sherman04} A.~Sherman,  Phys. Rev. B  {\bf 70}, 184512 (2004);
%( Evolution of the hole and spin-excitation spectra of the
%two-dimensional $t-J$ model: From light to heavy doping);
A.~Sherman  and M.~Schreiber,  Fiz. Nizk. Temp. (Low Temp. Phys.,
Ukraine)  32,  499 (2006).
%Spin dynamics in cuprate perovskites.
\bibitem{Sega03}  I. Sega,  P. Prelov\v{s}ek, and J. Bon\v{c}a,
  Phys. Rev. B  {\bf 68}, 054524, (2003).
% Magnetic fluctuations and resonant peak in cuprates:
% Towards a microscopic theory.
\bibitem{Prelovsek04} P. Prelov\v{s}ek, I. Sega, and J. Bon\v{c}a,
Phys. Rev. Lett. {\bf  92}, 027002 (2004).
%Scaling of the  Magnetic response in doped antiferromagnets.
\bibitem{Sherman03a} A. Sherman and M. Schreiber,
Phys. Rev. B  {\bf 68}, 094519 (2003).
%Resonance peak in underdoped cuprates.
\bibitem{Sega06}    I. Sega  and  P. Prelov\v{s}ek
Phys. Rev. B {\bf 73}, 092516 (2006).
%Double dispersion of the magnetic resonant mode in cuprates from
%the memory function approach.
\bibitem{Prelovsek06}  P. Prelov\v{s}ek and I. Sega,
Phys. Rev. B  {\bf 74}, 214501 (2006).
%Magnetic collective mode in underdoped cuprates: a
% phenomenological analysis.
\bibitem{Vladimirov05} A.A. Vladimirov, D. Ihle, and N. M. Plakida,
Theor. Math. Phys. {\bf 145}, 1576 (2005).
%Theoretical and Mathematical Physics, 145(2): 1575–1588 (2005)
%Dynamical spin susceptibility in the t–J model: the
%memory function method.
\bibitem{Winterfeldt98} S. Winterfeldt and D. Ihle.
Phys. Rev. B {\bf 58}, 9402  (1998).
% Theory of magnetic short-range order in the t-J model.
\bibitem{Winterfeldt97}  S. Winterfeldt and D. Ihle,
Phys. Rev. B {\bf 56}, 5535 (1997).
% Theory of antiferromagnetic short-range order in the
%two-dimensional Heisenberg model.
\bibitem{Winterfeldt99}  S. Winterfeldt and D. Ihle.\
Phys. Rev. B {\bf 59}, 6010 (1999).
% Spin dynamics and antiferromagnetic short-range order in the two-dimensional
%Heisenberg model.
\bibitem{Shimahara91} H. Shimahara and S. Takada,
J. Phys. Soc. Jpn. {\bf 60}, 2394 (1991).
\bibitem{Jackeli99}  G. Jackeli and N. M. Plakida, Phys. Rev. B
 {\bf 60}, 5266 (1999).
\bibitem{Bonca89} J. Bon\v{c}a, P. Prelov\v{s}ek,
and I. Sega, Europhys. Lett. {\bf 10}, 87 (1989).
\bibitem{Wiese94} U.-J. Wiese and H.-P. Ying, Z. Phys. B
{\bf 93}, 147 (1994).
\bibitem{Vojta96} M. Vojta and K. Becker, Phys. Rev. B
 {\bf 54}, 15483 (1996).
\bibitem{Jaklic96}  J. Jakli\v{c} and P. Prelov\v{s}ek,
Phys. Rev. Lett.  {\bf 77}, 892 (1996).
%  Thermodynamic properties of the planar $t-J$ model.
\bibitem{Shimahara92} H. Shimahara and S. Takada,
J. Phys. Soc. Jpn. {\bf 61}, 989 (1992).
\bibitem{Torrance89}
J.B. Torrance, A. Bezinge, A.I.~Nazzal, T.C.~Huang,
S.S.P.~Parkin, D.T.~Keane, S.J.~LaPlaca,
 P.M.~Horn, and G.A.~Held,  Phys. Rev. B  {\bf 40}, 8872 (1989).
% Properties that change as superconductivity disappears at high-doping
%concentrations in La2-xSrxCuO4.
\bibitem{Zavidonov98}  A. Yu. Zavidonov and D. Brinkmann,
Phys. Rev. B   {\bf 58}, 12486 (1998).
\bibitem{Tyc90} S. Ty\v{c} and B. Halperin, Phys.~Rev.~B {\bf 42}, 2096
(1990).
\bibitem{Makivic92} M. Makivi\'{c} and M. Jarrell, Phys. Rev. Lett.
 {\bf 68}, 1770 (1992).
\bibitem{Ihle94}  D. Ihle and N. M. Plakida,
Z. Phys. B  {\bf 96}, 159 (1994).
\bibitem{Keimer92} B. Keimer, N.~Belk, R.J.~Birgeneau,
A.~Cassanho, C.Y.~Chen,  M.~Greven,  M.A.~Kastner, A.~Aharony,
Y.~Endoh, R.W.~Erwin, and G.~Shirane, Phys.~Rev.~B {\bf 46},
14034 (1992).
\bibitem{Stock08} C. Stock,  W.J.L.~Buyers, Z.~Yamani,
Z.~Tun,  R.J.~Birgeneau, R.~Liang, D.~Bonn, and W.N.~Hardy,
 Phys.~Rev.~B {\bf 77}, 104513 (2008).
 %Magnetic excitations in pure,lightly doped, and weakly metallic
%La$_{2}$CuO$_4$.
\bibitem{Kakurai93} K. Kakurai, S.~Shamoto, T.~Kiyokura, M.~Sato, J.M.~Tranquada,
and G.~Shirane,  Phys.~Rev.~B {\bf 48}, 3485 (1993).
\bibitem{Plakida99}  N.M.~Plakida   and   V.S. Oudovenko,
 Phys.~Rev.~B {\bf 59}, 11949 (1999).
%Electron spectrum and superconductivity in the t-J model at
%moderate doping,
\bibitem{Vladimirov07} A.A. Vladimirov, D. Ihle, and N. M. Plakida,
Theor. Math. Phys. {\bf 152}, 1331 (2007).
%Static spin susceptibility in the t–J model

\end{thebibliography}
\end{document}